\documentclass[useAMS,usenatbib]{mn2e}
\usepackage{epsfig, bm, times, aas_macros}
\usepackage[english]{babel}
\usepackage{amsmath}
\usepackage{amssymb}
\usepackage{graphicx}
\usepackage{color}

\newcommand{\tws}[1]{http://cococubed.asu.edu/code\_pages/#1}
\newcommand{\intza}{http://www.sron.nl/$\sim$jeanz/bursterlist.html}
\newcommand{\code}[1]{\texttt{#1}}
\newcommand{\timt}{\times}
\newcommand{\bc}{art-2012-braith-cave}
\newcommand{\nuke}{\rm{n}}
\newcommand{\kaco}{\kappa_{\rm{c}}}
\newcommand{\tent}[1]{\times 10^{#1}}
\newcommand{\dtempdt}[1]{T_{\rm t,#1}}
\newcommand{\ro}{R_{\rm Ro}}
\newcommand{\vf}{v_{\rm{f}}}%\newcommand{\vef}{v_{\rm{eff}}}
\newcommand{\scal}[3]{\left(\frac{#1}{#2}\right)^{#3}}
\newcommand{\ddif}{{\rm D}}
\newcommand{\derat}[1]{_{#1}}
\title[Flame Propagation during Type I X-ray Bursts] {Flame
  Propagation on the Surfaces of Rapidly Rotating Neutron Stars during
  Type I X-ray Bursts}

\author[ Y.\ Cavecchi, A. L.~Watts, J.\ Braithwaite, Y. Levin]{Yuri
  Cavecchi$^{1,2}$\thanks{E-mail: ycavecchi@uva.nl}, Anna
  L.~Watts$^{1}$, Jonathan Braithwaite$^{3}$, Yuri
  Levin$^{4,2}$\\ $^{1}$Astronomical Institute ``Anton Pannekoek'',
  University of Amsterdam, Postbus 94249, 1090 GE Amsterdam, The
  Netherlands\\ $^{2}$Sterrewacht Leiden, University of Leiden, Niels
  Bohrweg 2, NL-2333 CA Leiden, The Netherlands\\ $^{3}$Argelander
  Institut f\"ur Astronomie, Universit\"at Bonn, Auf dem H\"ugel 71,
  53121 Bonn, Germany\\ $^{4}$Monash Center for Astrophysics and
  School of Physics, Monash University, Clayton, VIC 3800, Australia}

\pagerange{\pageref{firstpage}--\pageref{lastpage}} \pubyear{}

\begin{document}
\maketitle
\label{firstpage}

\begin{abstract}
% the abstract
  We present the first vertically resolved hydrodynamic simulations of
  a laterally propagating, deflagrating flame in the thin helium ocean
  of a rotating accreting neutron star. We use a new hydrodynamics
  solver tailored to deal with the large discrepancy in horizontal and
  vertical length scales typical of neutron star oceans, and which
  filters out sound waves that would otherwise limit our timesteps. We
  find that the flame moves horizontally with velocities of the order
  of $10^5$ cm s$^{-1}$, crossing the ocean in a few seconds, broadly
  consistent with the rise times of Type I X-ray bursts. We address
  the open question of what drives flame propagation, and find that
  heat is transported from burning to unburnt fuel by a combination of
  top-to-bottom conduction and mixing driven by a baroclinic
  instability. The speed of the flame propagation is therefore a
  sensitive function of the ocean conductivity and spin: we explore
  this dependence for an astrophysically relevant range of parameters
  and find that in general flame propagation is faster for slower
  rotation and higher conductivity.
\end{abstract}


\section{Introduction}
\label{sec:intro}

Type I bursts are tremendous thermonuclear explosions on the surface
of accreting neutron stars (NSs), with luminosities that can easily
reach the Eddington limit. They are characterized by a fast increase
in the X-ray luminosity, known as the rise, that lasts from less than
a few seconds up to tens of seconds, and by an exponential, slow
decay, the tail, that lasts from tens of seconds to a few minutes
\citep[]{rev-1993-lew-par-taa, rev-2008-gal-mun-hart-psal-chak}. More
than 90 known X-ray sources have shown Type I bursts (for an
up-to-date list see In't Zand web page \intza). All are low mass X-ray
binaries, where the NS accretes matter from the outer layers of the
companion.

Whether the accreted fluid spreads freely or is confined to part of
the NS surface probably depends on whether the accretion takes place
via a boundary layer \citep[the disc directly `touching' the
NS,][]{art-2010-inoga-suny} or via magnetic channelling and,
subsequently, on the strength of the magnetic field itself
\citep{art-1998-brow-bild}. The majority of bursters do not show
persistent pulsations, while those that do feature very weak magnetic
fields. It is therefore reasonable to assume that the accreted
material spreads over all of the NS surface, forming a thin highly
combustible ocean consisting of mostly light elements.

As new fluid piles up, the deeper layers are compressed until the
temperature and density are high enough to trigger nuclear reactions
of H, He or both. Depending on the accretion rate and the composition
of the fluid, the burning can be stable or unstable \citep[]{
  art-1981-fuj-han-miy}: in the latter case Type I bursts occur. It
seems unlikely that the whole star ignites at the very same moment
\citep{art-1982-shara}; instead it is more likely that the ocean
ignites locally and that the resulting flame propagates laterally and
engulfs the whole (or a substantial fraction) of the NS surface. It is
the physics of the lateral propagation of the thermonuclear flame that
is the focus of this study. Our goal is to investigate the open
question of what controls the propagation of the flame and the
ignition of unburnt fuel. Mechanisms that may be involved include
conduction, turbulent mixing associated with convection or other
hydrodynamical instabilities, or compression (of unburnt fuel by
elements that are already burning and hence expanding). Answering this
question is critical to explaining burst time scales, such as the rise
times. It is also relevant to the development of burst oscillations
(fluctuations in the intensity of the burst lightcurves, see
\citeauthor{rev-2012-watts} \citeyear{rev-2012-watts} for a
review). The mechanisms suggested to explain such oscillations always
involve some kind of asymmetry on the burning surface, such as the
presence of a hot-spot \citep{art-1996-stro-etal} or surface mode
patterns excited by the flame \citep{art-2004-hey,art-2005-piro-bild,
  art-2008-berk-levin}. \citet[hereafter
SLU]{art-2002-spit-levin-ush}\defcitealias{art-2002-spit-levin-ush}{SLU}
suggested that the Coriolis force might confine burning material in
`thermonuclear hurricanes', pointing out the importance of the
rotation of the star and of hydrodynamics for the flame propagation.

Early attempts at multi-dimensional simulations of the flame
propagation mechanism include \cite{art-1984-noza-ikeu}, who followed
a zonal approach, and \cite{art-1982-fryx-woos-a}, who
hydrodynamically simulated the first few milliseconds of a detonation
in a thick layer.  \citet{art-2001-zing-etal} continued this line of
research, following the detonating flame for up to hundreds of
milliseconds. However the thick layers required for detonation are
unlikely to accumulate between bursts, so that in reality the flame
probably develops via deflagration \citep{art-2011-malo-etal}.  The
most recent studies of flame propagation, which last again only a few
milliseconds (far shorter than the timescales of real X-ray bursts),
are those by
\citet{art-2012-sim-gryaz-etal-a,art-2012-sim-gryaz-etal-b}.  In the
first of these papers, the flame propagates via a detonation wave, in
a manner similar to that in \citet{art-2001-zing-etal}.  The authors
find that propagation is due to the hot fluid expanding and spilling
over the top of the cold fluid, which then compresses and ignites.  In
the second paper, the authors explore regimes with densities $\sim
2\times 10^7 \hbox{g}/\hbox{cm}^3$ at the base of the ocean,
conditions where one might expect deflagration rather than
detonation. However, in this case the simulations did not develop a
steadily propagating flame.

None of these studies, however, took into account the rotation of the
NS. \citetalias{art-2002-spit-levin-ush} studied the role of the
Coriolis force in flame dynamics and concluded that the Rossby
adjustment radius (the horizontal length scale over which the Coriolis
force becomes effective in laterally confining the high or
low-pressure region) may determine the horizontal scale of the burning
front. They also pointed out the importance of the ageostrophic
flow\footnote{A geostrophic flow is a general configuration in which
  the pressure forces are exactly balanced by the Coriolis force:
  ${\bm\nabla}P = 2\rho {\bf\Omega}\times{\bf u}$. The fluid motion is
  along the surfaces of constant pressure
  \citep[see][]{book-1987-Pedlo}. In an ageostrophic flow this
    condition does not hold.} at the hot-cold fluid interface as part
of the mechanism for the flame propagation itself. That said,
\citetalias{art-2002-spit-levin-ush} used a two layer, shallow water,
method to simulate the propagation of the flame and therefore had to
make phenomenological assumptions about heat and momentum transport in
the vertical direction. Our simulations do not involve such
assumptions, and are resolved in the vertical direction, allowing us
to make a detailed study of the flame-propagation physics, taking full
account of rotation.

In this paper, we simulate flame propagation on a domain that has a
horizontal extent that is a substantial fraction of the surface of the
NS. At the heart of our simulations is the hydrodynamical code
described in \citet[hereafter BC]{\bc}\defcitealias{\bc}{BC}. It is a
multidimensional code, which, by construction, enforces hydrostatic
equilibrium in the vertical direction (this assumption is justified
since the timescale for sound propagation in the vertical direction in
the burning layer is much shorter than the nuclear reaction
timescale). Hydrostatic equilibrium allows us to use a longer time
step, which would otherwise be limited by sound wave propagation in
the vertical direction.

There are other methods that remove sound waves: adopting an implicit
scheme is one efficient way, or one can use the basic constant-density
incompressible approximation (in which sound and buoyancy waves are
both absent), the anelastic approximation \citep{art-1962-ogura-phil}
or the Boussinesq approximation (see \citeauthor{art-1996-lilly}
\citeyear{art-1996-lilly} for a review). These latter approximations
can be used under the assumptions that the thermodynamic variables are
close to a hydrostatically balanced background state, that the
frequency of the motions is much less than the frequency of sound
waves and that the vertical to horizontal length scale ratio or the
motion is not too large. However, the hydrostatic approximation also
allows us to use \emph{pressure as the vertical coordinate}. This is a
great advantage, because we can follow the inflation of the fluid
without the need for extra grid points which would lie unused for most
of the simulation, consuming extra memory and increasing calculation
time \citepalias[for more details see][]{\bc}.

In Sec. \ref{sec:techpart} we briefly review the numerical code
described in \citetalias{\bc} and introduce the additions and
modifications made to the code in order to study thermonuclear flame
propagation. We then report our results on flame spreading in
Sec. \ref{sec:results}. We focus in particular on the mechanisms that
drive flame propagation and investigate the speed dependence on the
rotation rate and conductivity. We conclude with a brief summary in
Sec.  \ref{sec:conclusions}.

\section{Numerical implementation}
\label{sec:techpart}

In this section we describe the modifications made to the code
reported by \citetalias{\bc} in order to make it suitable for study of
flame propagation during Type I X-ray bursts.  First however we
briefly review the most salient features of the code as outlined in
\citetalias{\bc}.  It is a 3D magnetohydrodynamical code, which uses
the $\sigma$-coordinate system (a pressure coordinate system, see
below) on a staggered grid: thermodynamical variables such as
temperature, pressure, density and heat sources are evaluated at the
centres of the grid cells, while velocities are evaluated on the
``faces'' of the cells. The code is 3D, but for this paper we use a 2D
version, assuming that variables are independent of one of the
horizontal dimensions.  We also neglect magnetic fields, postponing
this for future research.

The assumption of vertical hydrostatic equilibrium, justified by the
short vertical sound crossing time (much shorter than any time scale
of interest in burst simulations), allows us to discard vertically
propagating sound waves, numerical resolution of which consumed
the lion's share of the CPU time in previous numerical experiments.
In our simulations we can therefore employ much longer timesteps than
previous studies. Vertical equilibrium leads naturally to the
introduction of a vertical pressure coordinate. This in turn makes it
possible to follow the fluid as it expands, without the need for extra
grid cells.

The code evolves the two horizontal components of the fluid velocity,
the pressure (which acts as a pseudo density) and the temperature
using a three step Runge-Kutta scheme, while the spatial derivatives
are calculated with sixth-order finite differences.  Pressure is
defined as
\begin{equation}
\label{eq:Pdef}
P = \sigma P_* + P_{\rm{T}}
\end{equation}
where $P_*=P_{\rm B} - P_{\rm{T}}$ and $P_{\rm B}$ and $P_{\rm{T}}$
are the pressure at the bottom and at the top of the simulation.
$P_{\rm{T}}$ is a constant parameter in the simulations and
$\sigma\in[0,1]$ becomes the vertical coordinate ($0$ corresponding
to the top and $1$ to the bottom). $P_*$ can be shown to become a
pseudo density and the continuity equation becomes
\begin{equation}
\label{eq:Pevol}
\frac{\partial P_\ast}{\partial t} =- I_{\sigma=1}
\end{equation}
where
\begin{equation}
I\equiv \int^\sigma_0\!\!{\bm\nabla}_\sigma 
\!\cdot\!(P_\ast{\bf u}) \,{\rm d}\sigma^\prime,
\end{equation}
${\bf u}$ being the horizontal component velocity and
${\bm\nabla}_\sigma={\bf \hat{x}}
\partial_x + {\bf \hat{y}} \partial_y$, taken at constant $\sigma$.

Defining the Lagrangian derivative as $\ddif/\ddif t=\partial_t +
u_x\partial_x+ u_y\partial_y+\dot\sigma\partial_\sigma$, the momentum
and energy equations are
\begin{align}
  \frac{\ddif{\bf u}}{\ddif t} =& -{\bm\nabla}_\sigma \phi -
  \sigma\frac{{\bm\nabla}P_\ast}{\rho} - 2 {\bf\Omega}\times{\bf u} + {\bf F}_{\rm visc}.\\
  c_{\rm P} \frac{\ddif T}{\ddif t} =& \frac{1}{\rho}\frac{\ddif P}{\ddif t}
  + Q,
\end{align}
where 
\begin{equation}
\phi = gz = P_\ast \int_\sigma^1 \!\frac{\, {\rm d}\sigma^\prime}{\rho},
\end{equation}
$\rho$ is the density, ${\bf\Omega}$ is the rotation vector of the
star, parallel to the $z$ direction, ${\bf F}_{\rm visc}$ are the
viscous forces, $c_{\rm P}$ is the heat capacity at constant pressure
and $Q$ is the heat per unit mass per unit time (see \citetalias{\bc}
for more details); the equation of state (EOS) in \citetalias{\bc} is
a perfect monoatomic gas.

We now move on to discuss the modifications to the code that were
implemented to render it suitable for Type I burst simulations.

\subsection{Equation of state}
\label{sec:eos}

The first relevant change is to the EOS for the fluid in our
simulations, which has to be able to describe the physics of the NS
ocean. We take into account the composition of the fluid by expressing
it in terms of the mass fraction: X is the fractional mass of H, Y
that of He and $Z=1-X-Y$ the fraction of all heavier elements. For a
fully ionized perfect gas, the perfect gas EOS used in
\citetalias{\bc} becomes (assuming full ionization):
\begin{align}
P&=\frac{\rho R T}{\mu}\\
\intertext{with}
\mu&=\frac{12}{7 + 17X + 2Y}\,\rm{g\,mol}^{-1}\;.
\end{align}

However, since conditions in the NS ocean can lead to
electron degeneracy, which plays an important role in the vertical
support of the ocean against the gravitational field, we must take
this into account in our simulations. We still consider the atoms to
be fully ionized; however whilst the nuclei are assumed to be a
perfect gas, the electrons may be (partially) degenerate and
(partially) relativistic. We also need to include radiation pressure.

For this purpose we adapted the publicly available routine
\code{helmeos}\footnote{Available at \tws{eos.shtml}.} of
\citet{art-2000-tim-swes}. It uses the density $\rho$, temperature, X
and Y to derive pressure, energy, the thermodynamic potentials and
their derivatives with respect to $\rho$, T, X and Y.  In our code
structure $\rho$ is a derived quantity, while pressure is a primary quantity
\citepalias[see][sec. 2]{\bc}. To circumvent the problem of passing
density as an input parameter to the routine, we interface the original
\code{helmeos} with a zero-finding routine that calls it repeatedly
with different values of $\rho$ until convergence in pressure is
achieved.  Subsequent calls use the previous value of $\rho$ as an initial
guess. Given the Courant conditions we impose, the information in each
grid point does not change much in one time step, and convergence is
achieved within two or three calls. This is done in parallel for each
grid point.

The choice for the EOS is important for the evolution equation for the
temperature $T$.  We still derive it from the first law of
thermodynamics:
\begin{equation}
\frac{D E}{D t}=\frac{1}{\rho^2}P\frac{D \rho}{D t} + Q
\end{equation}
(where $E$ and $Q$ are the energy and heating rate per unit mass), but
we have two different results depending on the choice of the EOS. If
we use the perfect gas EOS, the evolution equation for temperature has
the form \citepalias[compare to][eq. 20]{\bc}:
\begin{equation}
\label{eq:tevolperfectgas}
\frac{c_P}{\mu}\frac{D T}{D t}= \frac{1}{\rho}\frac{D P}{D t} 
+ \frac{c_P T}{12}\left(17\frac{D X}{D t}+2\frac{D Y}{D t}\right)
+ Q,
\end{equation}
where $c_P=R(\gamma-1)/\gamma$, $\gamma$ is the adiabatic index and
$R=8.3144621\timt10^{7}$ erg K$^{-1}$ mol$^{-1}$ is the gas
constant\footnote{Note that in \citetalias{\bc} we included $\mu$ in
  the definition of $R$.}.

When we include electron degeneracy and radiation pressure, by
contrast, we have

\begin{equation}
\tilde c_P\frac{D T}{D t}=A\frac{D P}{D t} + B \frac{D X}{D t} + C \frac{D Y}{D t} + Q,
\end{equation}
with
\begin{equation}
D=\left(\frac{P}{\rho^2} - \frac{\partial E}{\partial \rho}\derat{T,X,Y}\right)
\end{equation}
\begin{align}
A&=\phantom{-}D\frac{\partial \rho }{\partial P}\derat{T,X,Y}\\
\tilde c_P&=\phantom{-}
\left(\frac{\partial E}{\partial T}\derat{\rho,X,Y} -D \frac{\partial \rho }{\partial T}\derat{P,X,Y}\right)
\\
\label{eq:derEcomp1}
B&=-\left(
\frac{\partial E}{\partial X}\derat{\rho,T,Y} -D \frac{\partial \rho }{\partial X}\derat{P,T,Y}\right)\\
\label{eq:derEcomp2}
C&=-\left(
\frac{\partial E}{\partial Y}\derat{\rho,T,X} -D \frac{\partial \rho }{\partial Y}\derat{P,T,X}\right),
\end{align}
where we make use of the relations
\begin{align}
\frac{\partial \rho }{\partial P}\derat{T,X,Y}&=
\phantom{-}\frac{1}{\partial P / \partial \rho\;\derat{T,X,Y}}\\
\frac{\partial \rho }{\partial T}\derat{P,X,Y}&=
-\frac{\partial P / \partial T\;\derat{\rho,X,Y}}{\partial P / \partial \rho\;\derat{T,X,Y}}\\
\label{eq:derrhocomp1}
\frac{\partial \rho }{\partial X}\derat{P,T,Y}&=
-\frac{\partial P / \partial X\;\derat{\rho,T,Y}}{\partial P / \partial \rho\;\derat{T,X,Y}}\\
\label{eq:derrhocomp2}
\frac{\partial \rho }{\partial Y}\derat{P,T,X}&=
-\frac{\partial P / \partial Y\;\derat{\rho,T,X}}{\partial P / \partial \rho\;\derat{T,X,Y}}\;,
\end{align}
since the routine returns variables as functions of $\rho$, $T$, $X$ and $Y$.

A further complication comes from the fact that \code{helmeos}
actually uses $\bar A=12/(1 + 11X + 2Y)$ and $\bar Z=\bar A(1 + X)/2$,
(instead of $X$ and $Y$ directly) for $P$, and that the derivatives of
$E$ and $P$ are evaluated with respect to $T$, $\bar A$, $\bar Z$ and
$\rho$. Therefore, for equations \eqref{eq:derEcomp1},
\eqref{eq:derEcomp2}, \eqref{eq:derrhocomp1} and
\eqref{eq:derrhocomp2} we also need
\begin{align}
\frac{\partial}{\partial X}&=-\frac{\bar A^2}{12}\left(
11\frac{\partial}{\partial\bar A}+(5-Y)\frac{\partial}{\partial\bar Z}
\right)
\\
\frac{\partial}{\partial Y}&=-\frac{\bar A^2}{12}\left(
\phantom{1}2\frac{\partial}{\partial\bar A}+(1+X)\frac{\partial}{\partial\bar Z}
\right)
\end{align}

The final evolution equation for the temperature is therefore
($D/Dt=\partial/\partial t + {\bf u}\cdot{\bm\nabla}_\sigma +
\dot{\sigma}\partial/\partial\sigma$):
\begin{equation}
\label{eq:progT}
  \partial T/ \partial t = \dtempdt{\rm{adv}} + \dtempdt{\rm{thermodyn}} + Q/\tilde c_P,
\end{equation}
where the contributions to $\partial T/ \partial t$ are separated into
\begin{align}
\dtempdt{\rm{adv}}&=-\left({\bf u}\cdot{\bm\nabla}_\sigma T +
\dot{\sigma}\partial T/\partial\sigma\right)
\\
\label{eq:thermodyn}
\dtempdt{\rm{thermodyn}}&=\frac{A}{\tilde c_P}\frac{D P}{D t} + \dtempdt{\mu}\\
\label{eq:thermodynmu}
\dtempdt{\mu}&=\frac{B}{\tilde c_P}\frac{D X}{D t} + 
\frac{C}{\tilde c_P}\frac{D Y}{D t}
\end{align}
and $Q/\tilde c_P$, so that we can test the relative importance of the
contributions of the different terms (see
Sec. \ref{sec:frontmecha}). $Q$ is further divided into $Q=Q_{\rm n} +
Q_{\rm cond} + Q_{\rm cool} + Q_{\rm hyper}$, where $Q_{\rm n}$ is the
nuclear burning contribution, $Q_{\rm cond}$ is the conduction
contribution and $Q_{\rm cool}$ is the cooling contribution from the
top (see the next sections). We also include an artificial diffusive
term $Q_{\rm hyper}$ (with a small coefficient, see Sec. 3.4 of
\citetalias{\bc}) to ensure numerical stability. In the case of the
perfect gas EOS the evolution equation is very similar to equation
\eqref{eq:progT}.

Finally, the term $D P/D t$ is evaluated according to equation (19) of
\citetalias{\bc}, which does not depend on the EOS, but on the choice
of the $\sigma$-coordinate system. $D X/D t$ and $D Y/D t$ have to be
treated carefully: in the case of reactions, or any change in
composition, they have to be evaluated explicitly (see
Sec. \ref{sec:burn-coll}).

\subsection{Conduction}
\label{sec:conduction}

Since conduction may play an important role in flame propagation, we
include a physical conduction term in $Q$ of the form:
\begin{equation}
Q_{\rm cond}=\frac{1}{\rho}\nabla\cdot\left(\frac{16 \sigma_{\rm{B}} T^3}
{3\rho \kaco} \nabla T \right)
\label{eq:phycond}
\end{equation}
where $\kaco$ is the effective opacity due to both radiative and
conductive processes.  In the $\sigma$-coordinate system $Q_{\rm
  cond}$ takes the form:
\begin{multline}
\frac{1}{P_\star}\left\{{\bm\nabla}_\sigma
  \left[
    \frac{16\sigma_BT^3}{3\kaco\rho}
    \left(
      \frac{P_\star}{\rho}{\bm\nabla}_\sigma T + 
      {\bm\nabla}_\sigma\phi\frac{\partial T}{\partial \sigma}
    \right)
  \right]+
\right.\\
+\frac{\partial}{\partial\sigma}
\left[
  \frac{16\sigma_BT^3}{3\kaco P_\star}
  \left(
    {\bm\nabla}_\sigma \phi^2\frac{\partial T}{\partial \sigma} 
    +\frac{P_\star}{\rho}
    {\bm\nabla}_\sigma \phi {\bm\nabla}_\sigma T
  \right.+
\right.\\
\left.
  \left.
    \left.      
      + g^2\frac{\partial T}{\partial\sigma}
    \right)
  \right]
\right\}
\label{eq:phycondfull}
\end{multline}
where $\phi= g z = P_\ast \int_\sigma^1 \!1/\rho\, {\rm
  d}\sigma^\prime.$ The terms that include $\phi$ are due to the fact
that the transformation matrix between the Cartesian coordinate system
and the $\sigma$ coordinate system is not everywhere orthogonal. These
contributions turn out to be small and can be neglected.

We have implemented two possibilities for evaluating $\kaco$: either
it has a fixed value set at the beginning of the simulation, or it is
calculated for each and every grid point taking into account the
composition and thermodynamical variables at that position. For
this second option we adapted the publicly available routines of
Timmes: \code{sig99}\footnote{Available at \tws{kap.shtml}.}. The
opacities calculated in this way take into account radiation,
scattering and the degree of degeneracy \citep[see][and references
therein]{art-2000-tim}. Based on the values of density, temperature
and composition in the simulations that we wanted to perform, we
decided to use an average constant value of $\kaco=0.07$ cm$^2$
g$^{-1}$ in our reference simulation, which speeds up the
calculations whilst still preserving the critical physics.

As anticipated in section \ref{sec:eos}, we include in equation
\eqref{eq:progT} a hyperdiffusive term (see \citetalias{\bc}, section
3.4.2), which mimics conduction. This term is unphysical and only used
to ensure numerical stability. In the horizontal direction, in
particular, it will be unphysically high, and may partly limit the
conclusions we can draw from our simulations. However, test runs
involving much lower hyperdiffusivity yielded flame velocities (see
section \ref{sec:anatomy}) which differ by only a few percent from the
values reported in this
paper. 

\subsection{Sources and sinks of heat: nuclear burning and cooling}
\label{sec:burn-coll}

Since we are interested in simulating Type I bursts, we implement
helium burning via the triple-$\alpha$ reaction according to
\citep[see][]{book-1984-clayton}:
\begin{equation}
\label{eq:nukeburning}
Q_{\nuke}=5.3\timt10^{18}\rho_5^2 \left(\frac{Y}{T_9}\right)^3 
e^{-4.4/T_9}\; \textrm{erg g$^{-1}$ s$^{-1}$},
\end{equation}
where $T_9$ is the temperature in units of $10^9$ K, $Y$ is the mass
fraction of He and $\rho_5$ is the density in units of $10^5$ g
cm$^{-3}$.  Including only the triple-$\alpha$ process is of course a
simplification, since there are many other reaction chains that should
be taken into account (this model would not be correct even for a pure
He accretor).  However we leave this refinement for later
investigation.

During burning the composition is evolved according to
\begin{align}
\label{eq:Yprogtot}
  \frac{DY}{Dt}&= - \frac{Q_{\nuke}}{\epsilon_\alpha}\\
  \intertext{which corresponds to}
\label{eq:Yprog}
\frac{\partial Y}{\partial t}&= -{\bf u}\cdot{\bm\nabla}_\sigma Y -
  \dot{\sigma}\frac{\partial Y}{\partial\sigma} -
  \frac{Q_{\nuke}}{\epsilon_\alpha}
\end{align}
where the first two terms come from advection and the third is the
consumption of He due to nuclear reactions
($\epsilon_\alpha=5.84\timt10^{17}$ erg g$^{-1}$ is the energy
production per gram per nucleon). We also include a form of artificial
diffusion as described in Sec. 3.4 of \citetalias{\bc} to ensure
numerical stability.

In terms of sinks of entropy, we include the possibility of cooling
from the uppermost layers. We use a simple formula, derived under the
assumption that energy is only transported through the layers above
the simulated computational domain, without additional
sinks or sources within the atmosphere. This is a somewhat coarse
approximation, particularly since expansion of the upper layers may
occur.  We use the temperature of the top grid cell in order to
evaluate the flux due to radiation and conduction:
\begin{equation}
F=\frac{16\sigma_B}{3\rho\kaco}T^3\frac{d T}{d z}
\end{equation}
where $F$ is the flux, which we assume to be constant in our plane
parallel approximation and $\sigma_B$ and $\kaco$ are as defined in
Sec. \ref{sec:conduction}.  We further assume that $\kaco$ is constant
in the layers above the simulation. Rearranging and integrating in
the vertical, $z$, direction, from the top of the simulation (${\rm
  T}$) to the top of the NS atmosphere (${\rm atm}$), we have
\begin{align}
F \int_{\rm atm}^{\rm T} \rho\,{\rm d}z &= 
\frac{16\sigma_B}{3\kaco}\int_{\rm atm}^{\rm T} T^3\,{\rm d}T\;.\\
\intertext{In hydrostatic equilibrium, the integral on the left hand side reduces to $P_{\rm T}/g$, so that}
F \frac{P_{\rm T}}{g}&=
\frac{4\sigma_B}{3\kaco}T^4\big|^{\rm T}_{\rm atm}\\
\intertext{Then, assuming that the temperature at the top of the atmosphere 
is negligible with respect to that at the top of the simulation, we obtain}
F&=\frac{4\sigma_B}{3\kaco P_{\rm T}/g}T^4_{\rm T}
\end{align}
This is the flux from the surface of a grid cell at the top of the
simulation.  In order to derive the entropy loss per unit mass, we
multiply the flux by the surface area $S$ of the cell and divide by
the mass within it:
\begin{align}
Q_{\rm{cool}}&= F \frac{S}{\rho_{\rm T} S \Delta z_{\rm T}}\\
\intertext{so that ($\Delta z \sim H \sim P_{\rm T}/g\rho_{\rm T}$)}
\label{eq:coolings}
Q_{\rm{cool}}&= \frac{4  g^2 \sigma_B}
{3\kaco P^2_{\rm T}}
T^4_{\rm T}
\end{align}
This is the sink term we use in our simulations; it could also be used
as a first approximation to calculate the bolometric lightcurve of the
bursts.

\subsection{Tracer particles}
\label{sec:tracpart}

Finally, we add the capability to follow tracer particles.  These are
assigned initial positions uniformly distributed in the integration
domain and are evolved according to
\begin{align}
\frac{d x}{dt}&=u_{\rm x}(x,y,\sigma)\\
\frac{d y}{dt}&=u_{\rm y}(x,y,\sigma)\\
\frac{d \sigma}{dt}&=\dot\sigma(x,y,\sigma)
\end{align}
where $d/d t$ is the rate of change of the particle's position in
$\sigma$ coordinates.  Time evolution is the same as for all of the
other variables \citepalias[see][section 3.3]{\bc}, and the values of
the three components of the velocity at arbitrary points within each
grid cell are derived by means of bilinear interpolation
\citep{book-1992-num-rec} of the fluid velocity\footnote{We also tested
  higher order interpolation methods, but found no significant
  differences.}.

\section{Flame propagation simulations}
\label{sec:results}

In this section we describe the numerical setup used for all the
simulations and then provide a description of what we see in the runs.
Finally, we describe our interpretation of what drives the flame
propagation.

\subsection{Numerical setup}
\label{sec:numsetup}

We ran a series of simulations resolving both the horizontal $x$ and
vertical $z$ directions, assuming that the dynamical variables are
independent of the $y$ coordinate (making the simulations effectively
2D). The fixed initial conditions, common to all of our simulations,
are
\begin{align*}
P_{\rm{T}}&=10^{22}\;\textrm{erg cm}^{-3} & 
P_*&=(e^{1.7} - 1)\tent{22}\;\textrm{erg cm}^{-3}\\
X&=0        & Y&=1\\
\nu_1&=0.03 & \nu_2&=0.5\\
\intertext{and}
T_0&=10^8\;\rm{K}   & \delta T&=3.81\,\tent8\;\rm{K}
\end{align*}
where $P_{\rm{T}}$ and $P_*$ are the pressure at the top and the
difference between the bottom and top pressure
\citepalias[see][section 2]{\bc}. Note that whilst $P_{\rm{T}}$ is
constant, $P_*$ is a function of horizontal position and time, but not
of $\sigma$.  The choice of $P_*$ means that we simulate 1.7 scale
heights.  $\nu_1$ and $\nu_2$ are the kinetic diffusive coefficients
\citepalias[see section 3.4 of][]{\bc}. The corresponding coefficients
for the temperature and the composition fractions $X$ and $Y$ are
taken to be 1\% of these values.

We also use a common initial temperature perturbation in all
simulations.  We use a $z$-independent temperature profile, which
varies in the horizontal direction according to:
\begin{equation}
T = T_0 + \frac{\delta T }{1 + \exp[(x - 1.2\;\rm{km})/0.36\;\rm{km}]}
\end{equation}
This function ensures that the temperature profile is smooth enough
that it does not cause numerical issues; $1.2$ km corresponds to the
position where the temperature perturbation of the background $T_0$ is
half of its maximum, while $0.36$ km is approximately half the width
between where the perturbation is asymptotic to its maximum and where
it is asymptotic to its minimum ($0$ K).

We simulate a domain with a horizontal extent of $7.5$ km, which
allows more than sufficient room for the propagating flame to reach a
steady state. The initial conditions have a high temperature at one
end of the domain, so the flame ignites there and propagates towards
the other end. In some sense the point where ignition occurs can be
thought of as the eye of the cyclonic system\footnote{A cyclone is a
  system of circulating fluid where, at a given height, the pressure
  at the centre is lower than at the sides. The fluid is drawn in at
  the bottom and launched to the top from the centre.}. We use
symmetric boundary conditions in the vertical direction and reflective
conditions in the horizontal direction. In all simulations presented
here, we use horizontal and vertical resolutions of 240 and
90. Gravitational acceleration $g=2\tent{14}$ cm s$^{-2}$ and we use
the plane-parallel approximation and a constant Coriolis parameter
($f=2\Omega$), i.e. the $f$-plane approximation. The fluid is at rest
at the beginning, $U_x=0$ cm s$^{-1}$, and quickly adjusts to the
Rossby solution \citepalias[see][sec. 4.2]{\bc} before the flame
spreads.
 
Since we want to study the effects of different rotation frequencies
and the influence of conduction, we run a series of models employing
different values of the spin frequency $\Omega$ and the opacity
$\kaco$. The parameters for the simulations that we run are given in
Table \ref{tab:runs}.

\begin{table}
\centering
\begin{tabular}{|r|r|c|c|}
Run&$\nu\;(\textrm{Hz})$&$\kaco\;(\textrm{g cm}^{-2})$&
$\vf\;(\textrm{cm s}^{-1})$\\
\hline
 1& 450$\phantom{.5}$&$1\tent{+0}$&$ (1.33\pm0.03)\tent{5}$\\
 2& 450$\phantom{.5}$&$7\tent{-1}$&$ (1.43\pm0.02)\tent{5}$\\
 3& 450$\phantom{.5}$&$5\tent{-1}$&$ (1.52\pm0.02)\tent{5}$\\
 4& 450$\phantom{.5}$&$3\tent{-1}$&$ (1.67\pm0.03)\tent{5}$\\
 5& 450$\phantom{.5}$&$1\tent{-1}$&$ (1.91\pm0.04)\tent{5}$\\
 6& 450$\phantom{.5}$&$7\tent{-2}$&$ (2.01\pm0.05)\tent{5}$\\
 7& 450$\phantom{.5}$&$5\tent{-2}$&$ (2.03\pm0.05)\tent{5}$\\
 8& 450$\phantom{.5}$&$1\tent{-2}$&$ (1.99\pm0.02)\tent{5}$\\
 9& 450$\phantom{.5}$&$1\tent{-3}$&$ (1.98\pm0.03)\tent{5}$\\
10&  50$\phantom{.5}$&$7\tent{-2}$&$ (1.11\pm0.18)\tent{6}$\\
11& 112.5            &$7\tent{-2}$&$ (5.30\pm0.31)\tent{5}$\\
12& 225$\phantom{.5}$&$7\tent{-2}$&$ (3.04\pm0.10)\tent{5}$\\
13& 900$\phantom{.5}$&$7\tent{-2}$&$ (1.39\pm0.02)\tent{5}$\\
\hline
\end{tabular}
\caption{Values of the spin frequency $\nu = \Omega/2\pi$ and the opacity $\kaco$ 
  used in the different simulations. In the third column
  we report the velocity of the flame as measured from the simulations, 
  with errors derived from the least squares fit (see section \ref{sec:anatomy}).
  See Appendix \ref{sec:convtest} for a discussion about the convergence rate 
  of the code and values of the flame speed.}
\label{tab:runs}
\end{table}

To help us find out how the flame propagates, we use test
particles and follow what happens to these fluid elements before,
during, and after ignition conditions are met.  We place the test
particles homogeneously in our grid (note that they are not
homogeneous in space since we use a pressure coordinate system) such
that $x_{i,j}=i* \delta x / 200$ and $\sigma_{i,j}=j * 1/200$, $i,j\in
[1,200]$ . We also track what happens at three different points in the
atmosphere with fixed horizontal position. These points rise and
descend with time, having fixed values of $\sigma$ not $z$, so
this approach is not strictly speaking Eulerian. However it still
allows us to see what happens when the flame reaches a determined
distance from the ignition point.

\subsection{General description of the propagating flame}
\label{sec:anatomy}

\begin{figure*}
  \centering
  \includegraphics[width=0.45\textwidth]{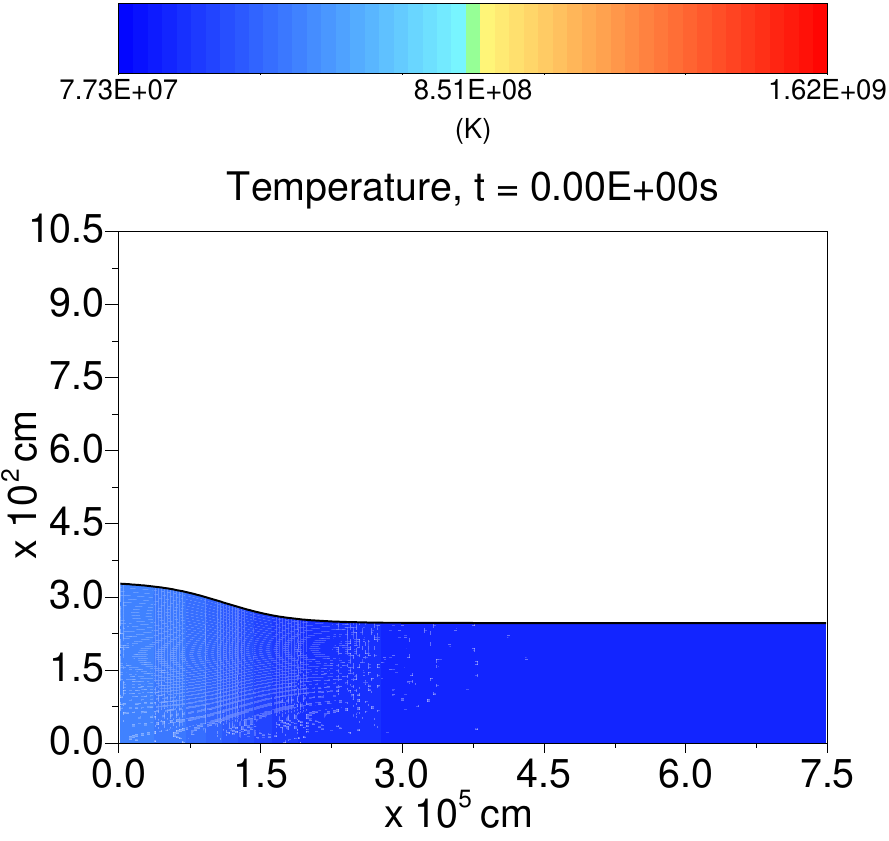}\hspace{\stretch{1}}
  \includegraphics[width=0.45\textwidth]{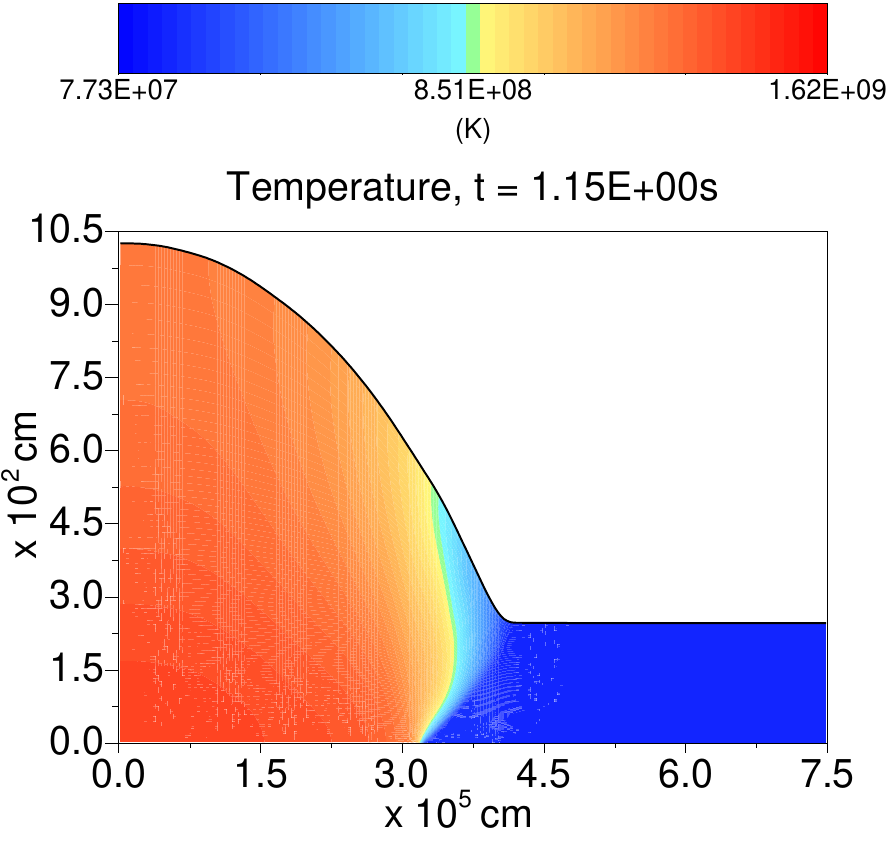}\\
  \includegraphics[width=0.45\textwidth]{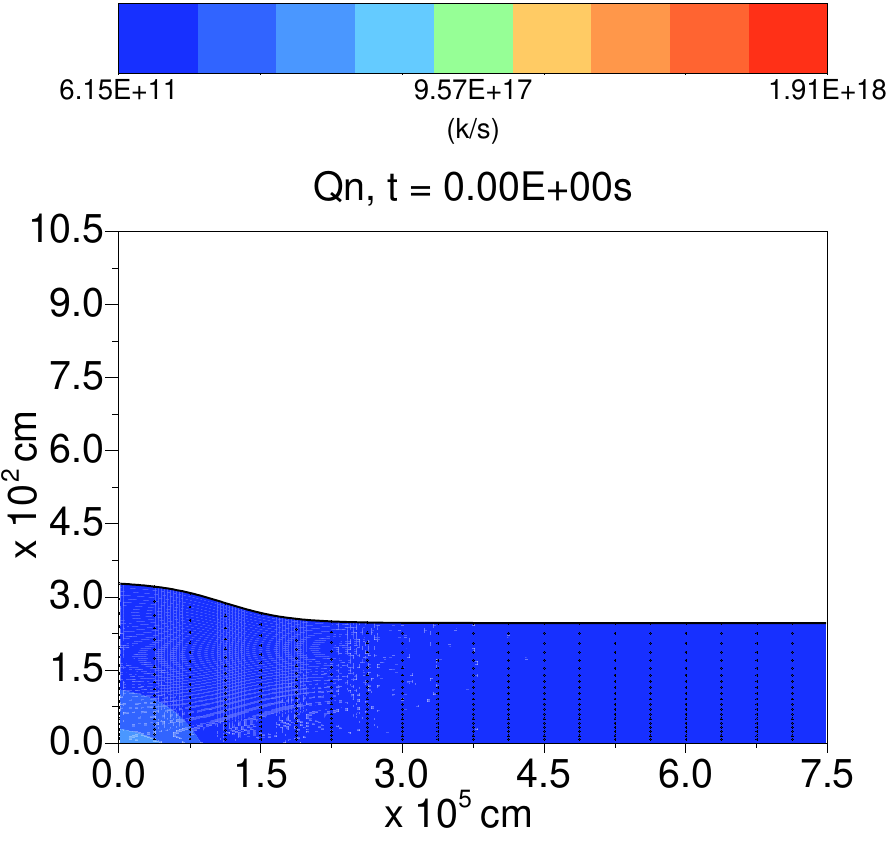}\hspace{\stretch{1}}
  \includegraphics[width=0.45\textwidth]{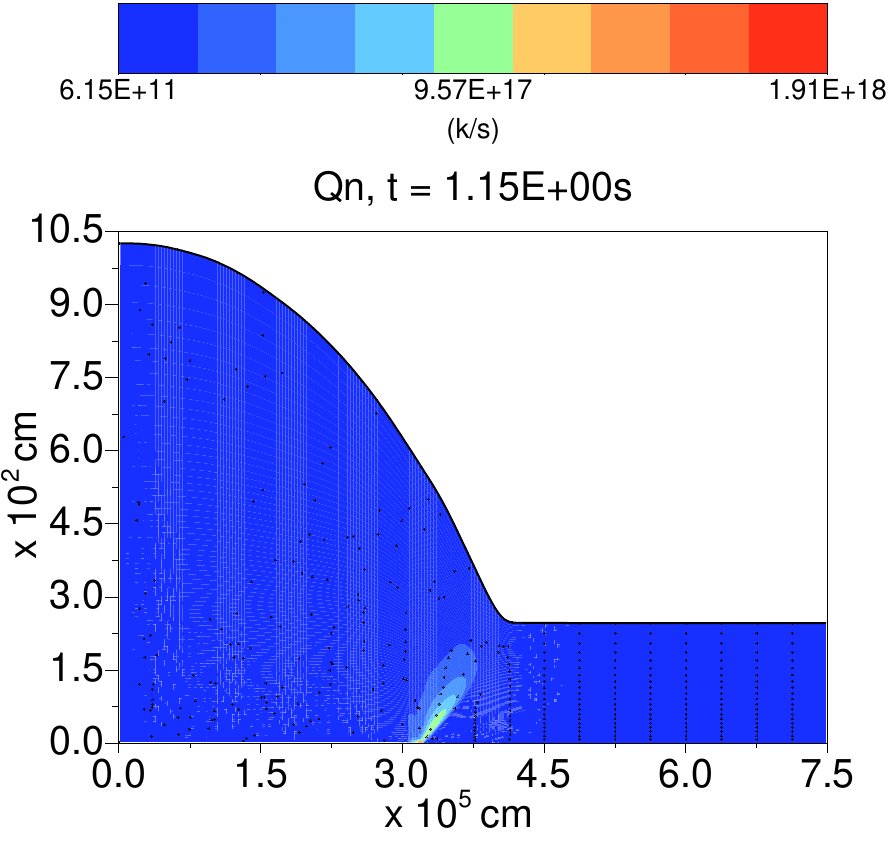}\\
  \includegraphics[width=0.45\textwidth]{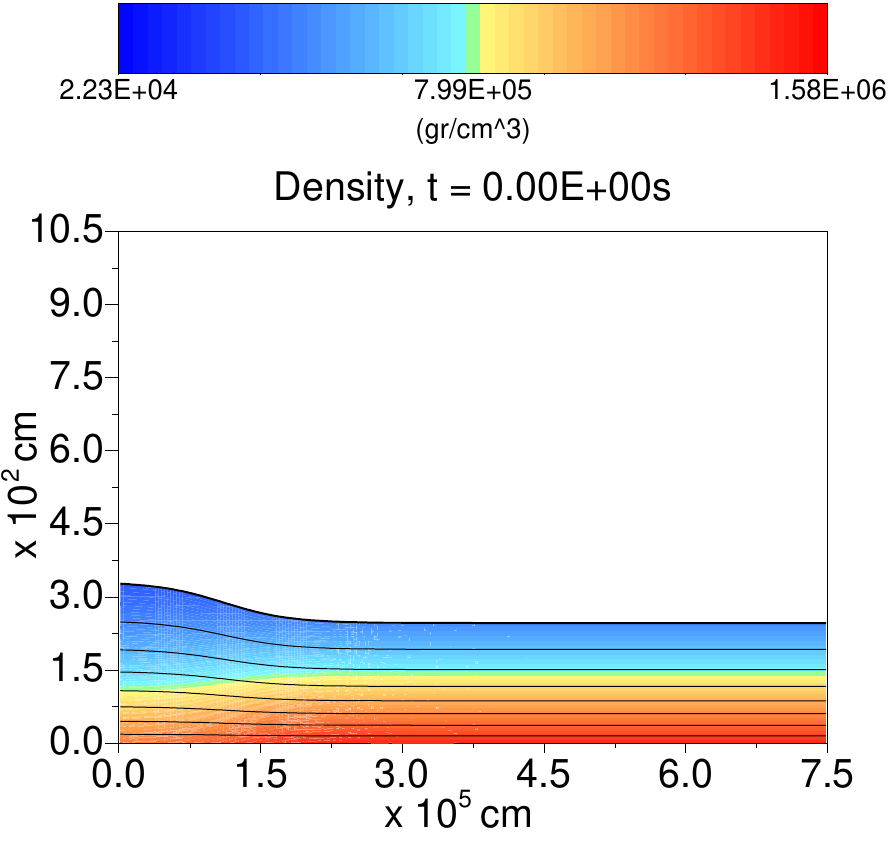}\hspace{\stretch{1}}
  \includegraphics[width=0.45\textwidth]{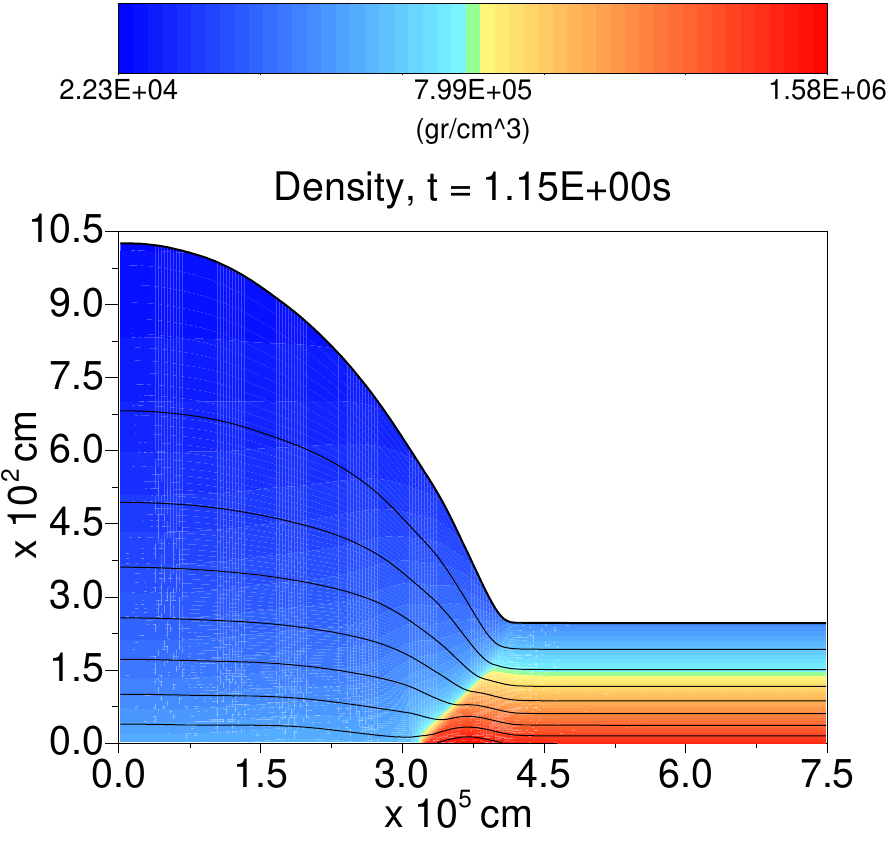}\\
  \caption{Initial conditions (left) and conditions at $t=1.15$ s
    (right), when the flame is steadily propagating, for reference
    simulation 6. The top panels show the temperature, the middle ones show
    burning rate with the tracer particles superimposed and the bottom
    panels show density with isobars superimposed (10 levels from
    $P=10^{22}$ to $6.2\tent{22}$ erg cm$^{-3}$). Note
    the different horizontal and vertical scales.}
\label{fig:anatomy1}
\end{figure*}

In this section, we give a qualitative description of the burning
fluid as a whole.  We begin by using one particular run as an example,
since the qualitative behaviour is general.  The left hand column of
Fig. \ref{fig:anatomy1} shows the fluid in its initial conditions
for reference run 6. The right hand column shows the conditions at
$t=1.15$ s, when the flame is propagating steadily. The upper panels
show the temperature distribution, and the middle ones the burning
rate. In these panels we superimpose our tracer particles. The bottom
panels show the density, with the iso--surfaces of pressure (isobars)
superimposed.
\begin{figure*}
  \centering
  \includegraphics[width=0.44\textwidth]{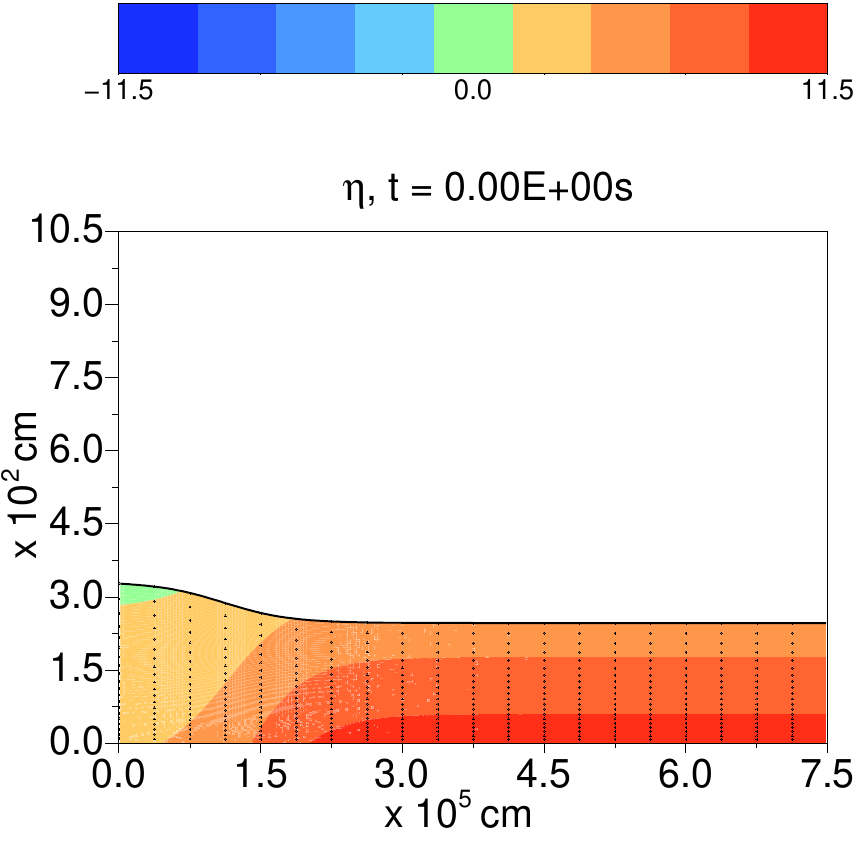}\hspace{\stretch{1}}
  \includegraphics[width=0.44\textwidth]{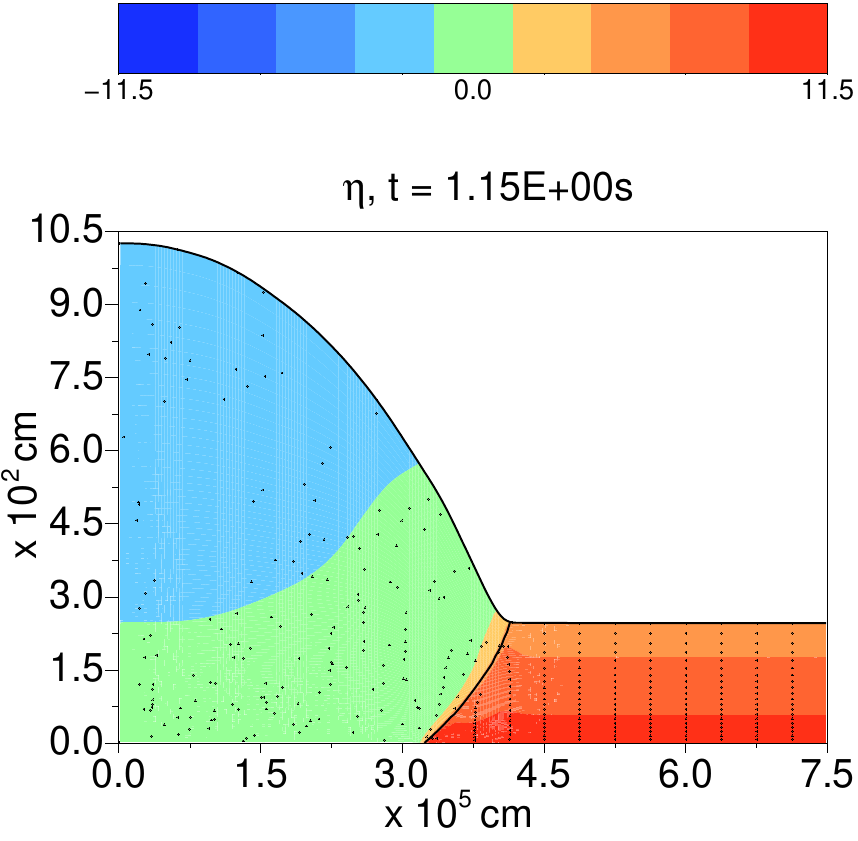}
  \caption{Electron chemical potential $\eta$ at the beginning and at
    $t=1.15$ s for reference run 6. Degeneracy decreases with lower
    $\eta$. The electrons are always partially degenerate, but
    degeneracy is partially lifted when the flame passes through. The
    black line again indicates the position of the interface.}
  \label{fig:eta}
\end{figure*}

In the top left panel of Fig. \ref{fig:anatomy1} the fluid is hotter
on the left of the image: this is the initial perturbation, where the
temperature is $T=4.81\tent{8}$ K, while at the other side the
temperature is $T=10^{8}$ K.  Moving to the right-hand panel, we see
that the flame front has moved to the right.  Where the fluid has
already burnt, it is hotter ($T\sim 10^9$ K) and has expanded by a
factor of the order of 4.\footnote{In general, the maximum expansion
  factor can be up to $\sim 4-5$ depending on the effective opacity
  $\kaco$ that sets the cooling rate.} Looking at the middle panels we
can see that the tracer particles have been scattered by the passage
of the front, while the lower panels show a drop in
density. Eventually, after the flame has passed (not shown in the
figure), the burning diminishes, the temperature decreases and the
fluid contracts.

Looking more closely at the propagating front, we see that it is
characterized by a slanted interface between the hot burning fluid on
the left, and the cold unburnt fluid on the right (in Fig.
\ref{fig:anatomy1} right, the interface lies roughly between
$x=3.7\tent{5}$ and $4.7\tent{5}$ cm). We see a decrease in
pressure on the left of the interface and an increase immediately to
the right of it (see the lowest isobars in the bottom right
panel of Fig. \ref{fig:anatomy1}). They reflect a change in
$P_\star$ (equation \ref{eq:Pdef}). Because of the hydrostatic
approximation, $P_\star$ is a measure of the column density at each
point. A change in $P_\star$ means horizontal mass motion. The
decrease before the front and the increase after it therefore show
that there has been a motion of matter from behind the front forward.

The electrons are partially degenerate, as can be seen from Fig.
\ref{fig:eta} where we plot the electron chemical potential. The
electrons remain partially degenerate throughout all the simulations,
but the degeneracy is lifted by the flame (as can be seen also by the
fact that the temperature increases by a factor $\sim 15$, while the
height of layer increases by only a factor of $4$ or $5$) so that in
the hot fluid the perfect gas pressure and the radiation pressure
become more important.

The peak of the burning is concentrated in a thin stripe along the
interface (Fig. \ref{fig:anatomy1}, middle right panel) where the
density is still high (undiminished by the increase in temperature)
and the fuel is still almost pure Helium. We also observe tracer
particles moving in the vertical plane, primarily along the interface
and in the region to the left of it.

\begin{figure*}
  \centering
  \includegraphics[width=0.44\textwidth]{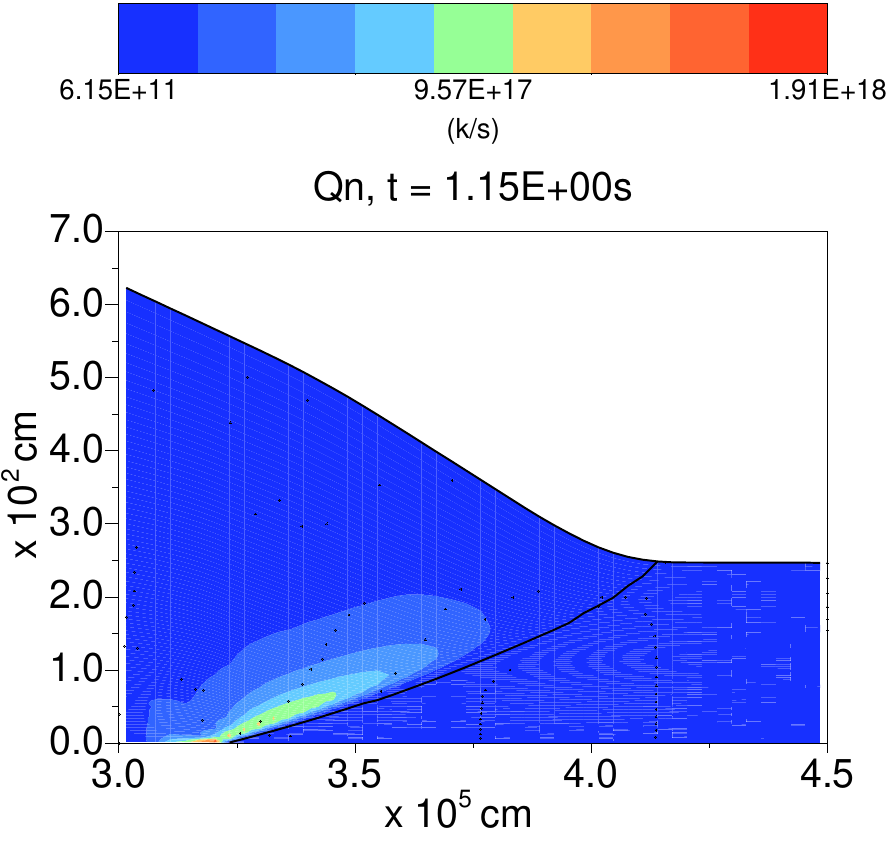}\hspace{\stretch{1}}
  \includegraphics[width=0.44\textwidth]{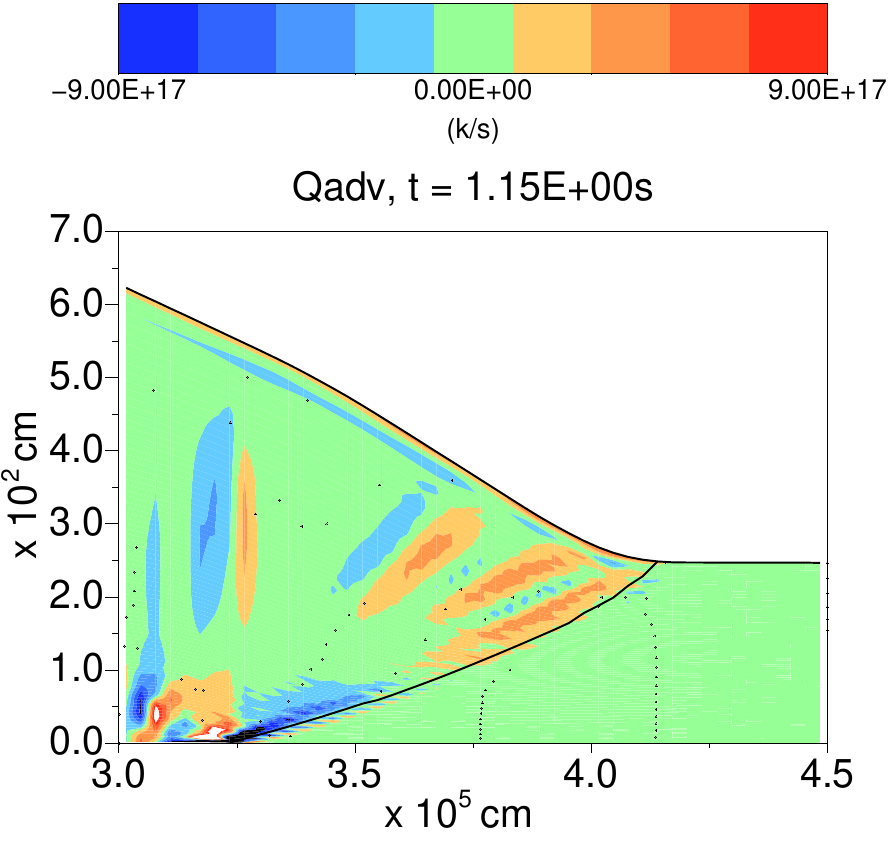}\\
  \includegraphics[width=0.44\textwidth]{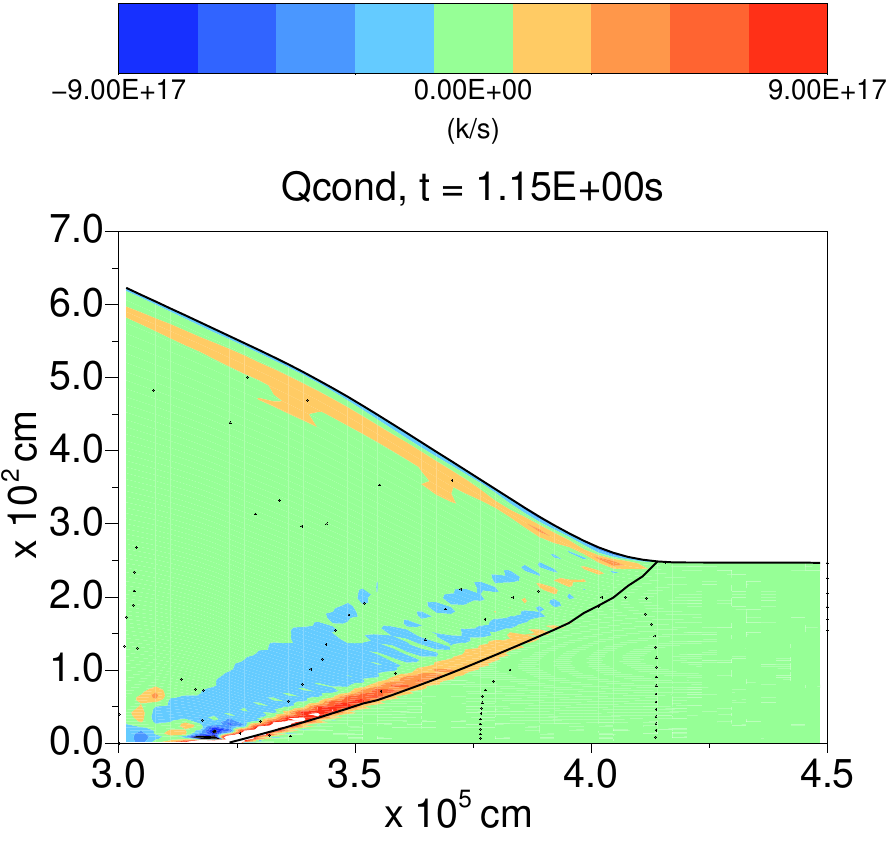}\hspace{\stretch{1}}
  \includegraphics[width=0.44\textwidth]{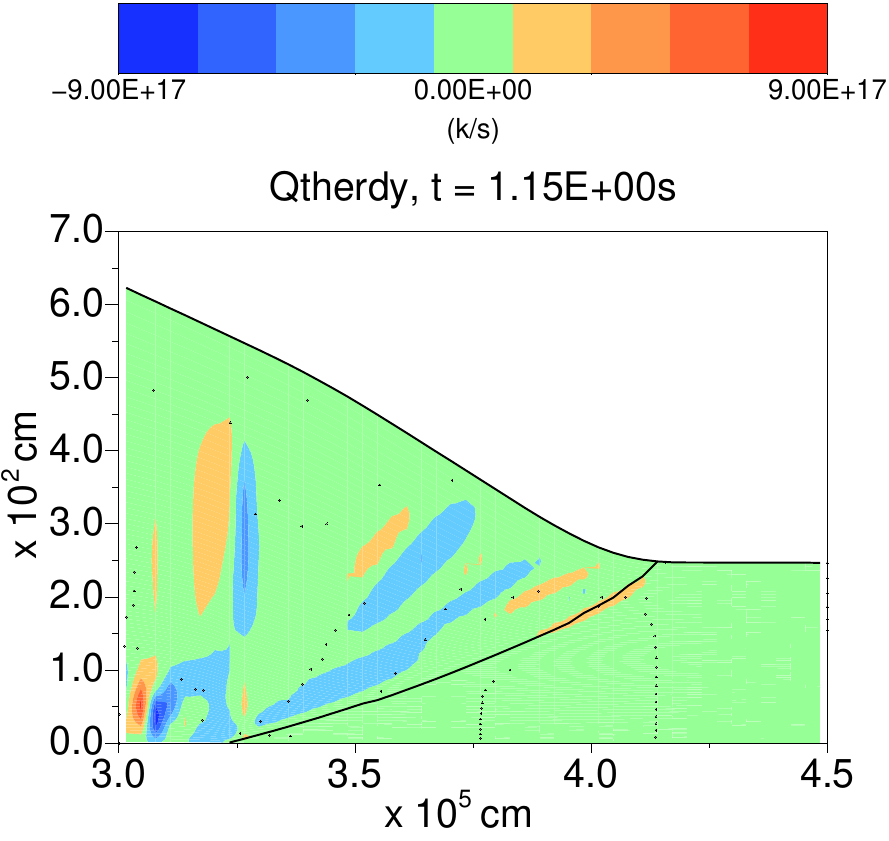}
  \caption{Burning rate and heating rate associated with advection
    $\dtempdt{adv}$, conduction $Q_{\rm{cond}}/\tilde c_P$ and
    thermodynamical compression $\dtempdt{thermodyn}$ for reference
    simulation 6. The black line indicates the hot-cold fluid
    interface. The colour scale has been restricted to highlight
    details (the white regions indicate values above the maximum of
    the scale and the black ones values below the minimum).}
  \label{fig:advcond}
\end{figure*}

In order to understand what is driving the flame forward, we measure
the different terms in the energy equation \eqref{eq:progT}:
conduction $Q_{\rm{cond}}/\tilde c_P$, advection $\dtempdt{adv}$
(motion of the fluid) and thermodynamic compression
$\dtempdt{thermodyn}$. For each term, we plot the contributions in
Fig. \ref{fig:advcond}. In the four panels the black line
approximates the interface between hot and cold fluid.  It is drawn
below the region of significant
burning in order to clearly demarcate regions where burning has
started from those where burning is about to start.

\begin{figure*}
  \centering
$\phantom{1}$\hspace{\stretch{1}}\includegraphics[width=0.425\textwidth]{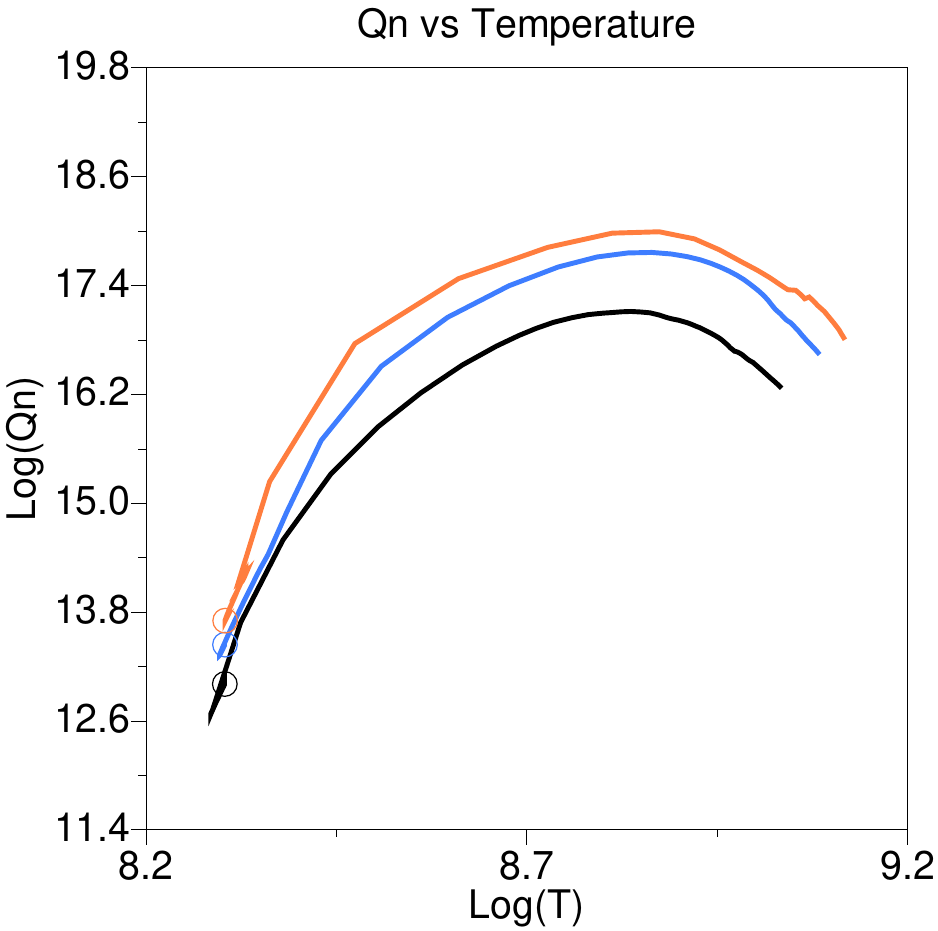}\hspace{\stretch{6}}
  \includegraphics[width=0.425\textwidth]{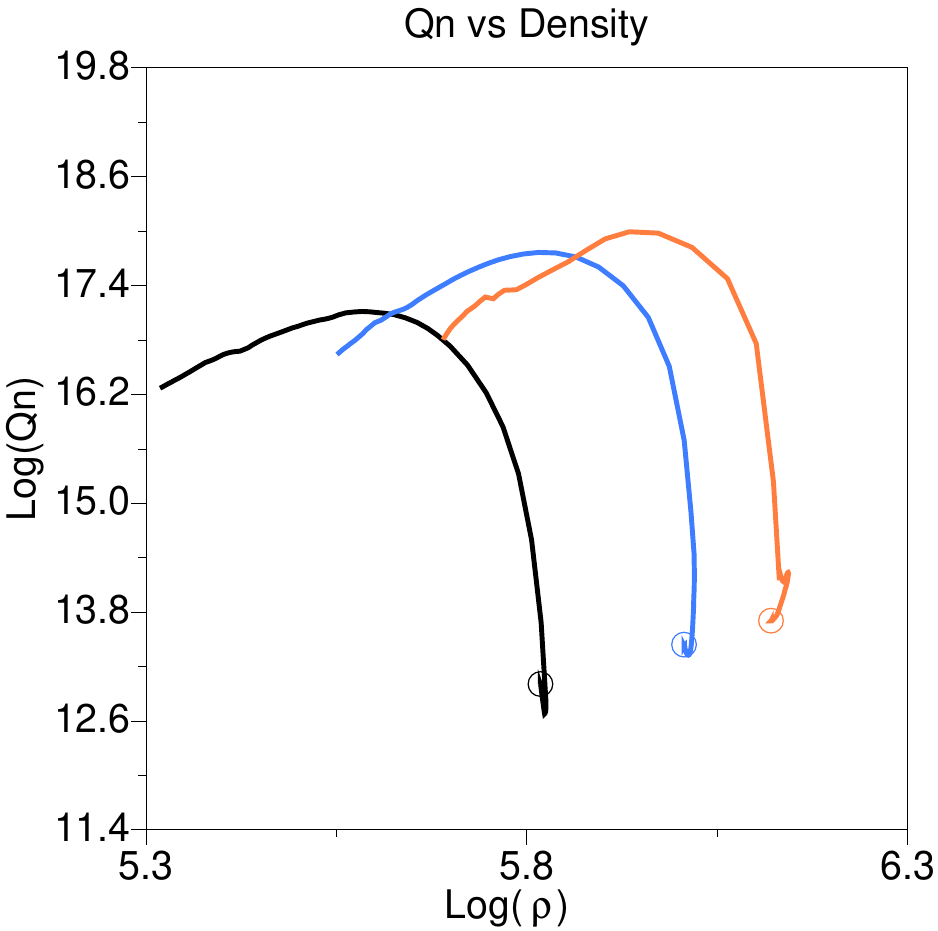}\hspace{\stretch{1}}$\phantom{1}$\\
$\phantom{1}$\hspace{\stretch{1}}\includegraphics[width=0.425\textwidth]{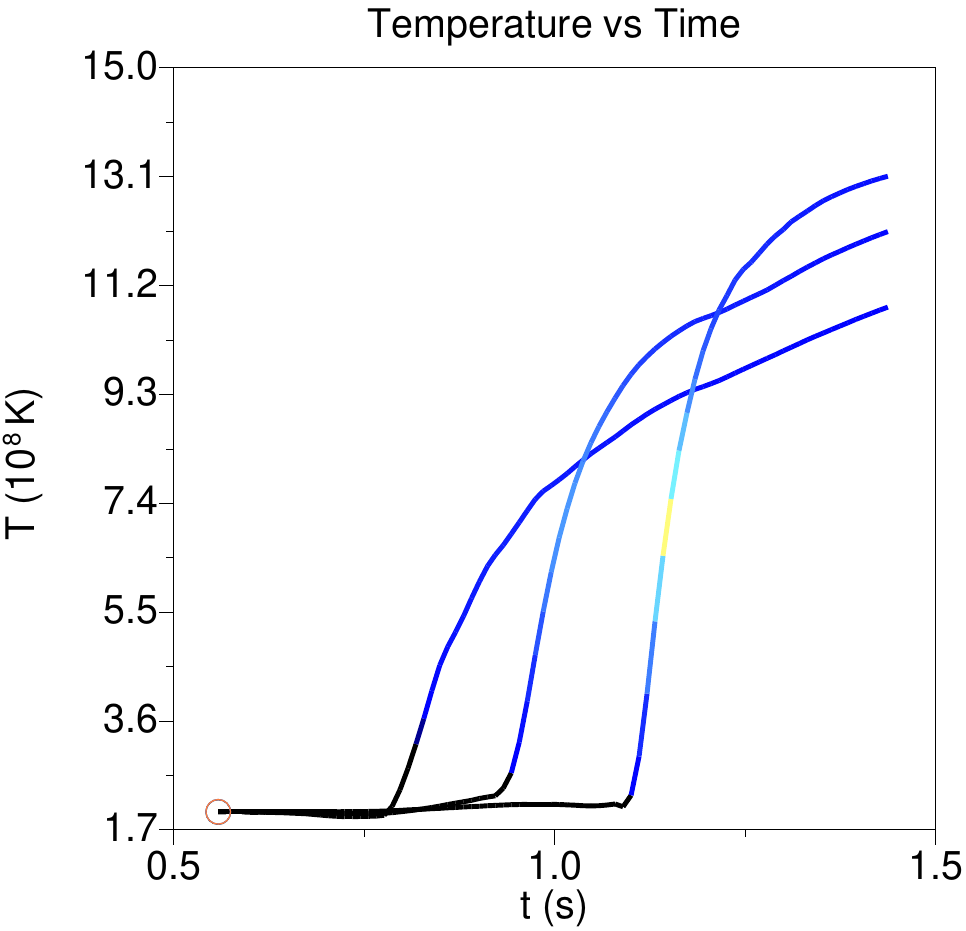}\hspace{\stretch{6}}
  \includegraphics[width=0.425\textwidth]{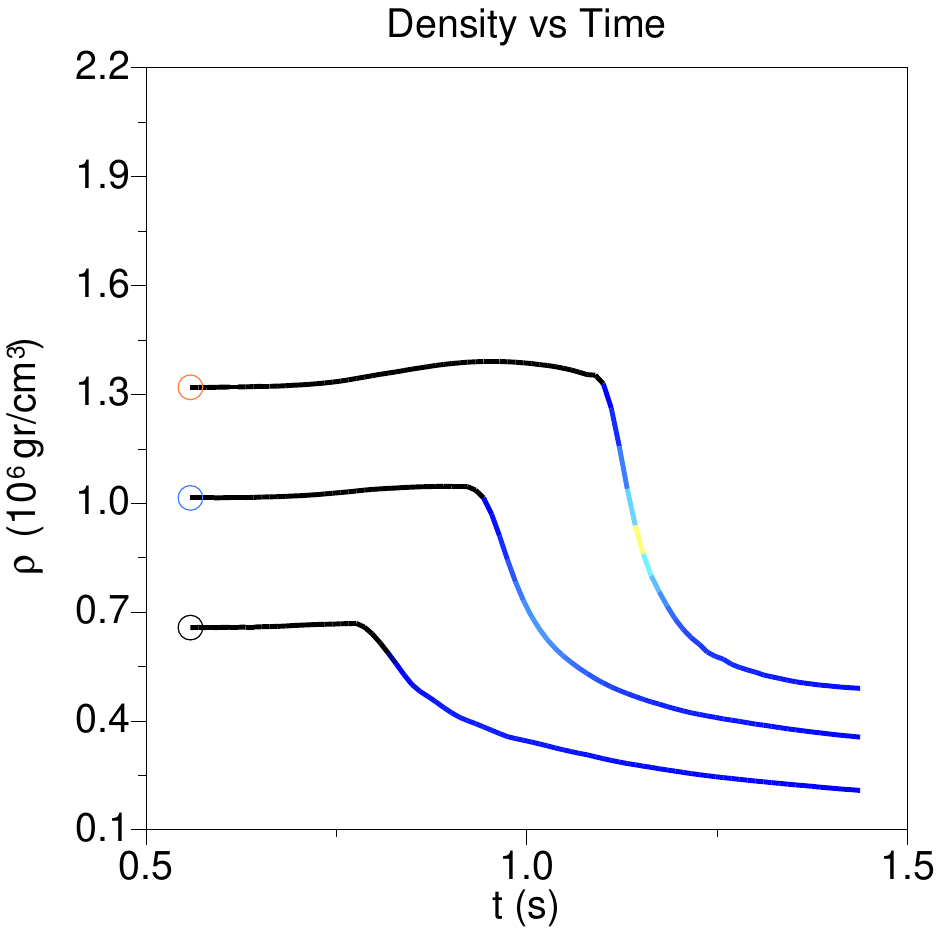}\hspace{\stretch{1}}$\phantom{1}$
  \caption{Top: burning rate versus temperature and density for three
    different points fixed relative to the grid, i.e. at fixed
    $x=3.3\tent{5}$ cm and sigma: near the top (black), in the middle
    (blue) and near the bottom (orange). Bottom: temperature and
    density versus time for the same points. The lines in the lower
    panels have colours corresponding to the same scale as the
    contours for the burning rate in the middle panel of Fig.
    \ref{fig:anatomy1}. The circles indicate the origin of the
    curves. The strong relation between ther burning rate and
    temperature is clear, while the importance of the change of
    density appears to be less.}
  \label{fig:fixpos}
\end{figure*}

It is clear from Fig. \ref{fig:advcond} that in the region
immediately below the peak of the burning, heat conduction is much
more important in increasing the temperature in the unburnt fuel
region than both the effects of mixing (measured by the advection of
temperature) or thermodynamic compression.  It is this process that
drives flame propagation, since the main burning occurs in this zone.
In the upper part of the interface, advection and thermodynamic
compression dominate heat transfer to the unburnt region. That picture
is confirmed by observing what happens at a fixed horizontal
position. In Fig. \ref{fig:fixpos} we plot the burning rate versus
temperature and density, and temperature and density versus time for
three different positions: at the top, in the middle and at the bottom
of the fluid at a fixed horizontal position $x=3.3\tent{5}$ cm. It can
be seen that the topmost point (black) in the figure is compressed and
its temperature rises.  The lower points then follow, but the burning
does not really start until the temperature has risen
sufficiently. The same figure also demonstrates how the burning rate
increases with increasing temperature, while the correlation with
density (see for example the lower panel) is not as strong. The
decrease of burning rate at the end of the curves is due to the
consumption of fuel which eventually becomes the most important
factor.

Directly \emph{above} the flame, on the other hand, heat conduction is
not effective, while the advective and thermodynamic compressive terms
show opposite signs. This is a clear signature of convection, which is
expected above the burning regions.  We note that the convective
cells near the topmost part of the hot-cold interface are mostly
parallel to it (i.e. almost horizontal, recall the extreme aspect
ratio of the interface), while the ones behind the interface are
\begin{figure}
  \centering
  \includegraphics[width=0.45\textwidth]{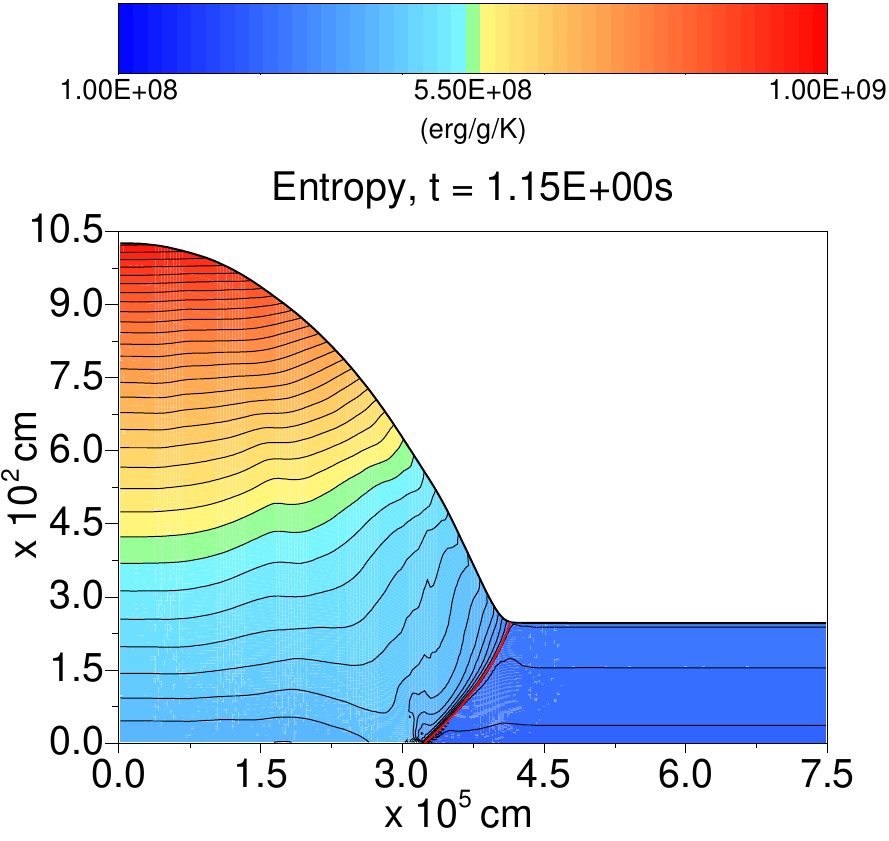}
  \caption{Entropy per unit mass (radiation, ion and electron gas) for
    reference simulation 6 at $t=1.15$ s. The black lines indicate
    the contours for better visualization. The red line indicates the
    position of the interface.}
  \label{fig:entropy}
\end{figure}
vertical\footnote{We want to stress that also these vertical cells are
  actually elongated in the horizontal direction due to the aspect
  ratio of our underlying grid cells.}. In Fig. \ref{fig:entropy} we
plot contours of the total entropy per unit mass as returned by the
code \code{helmeos}.

\begin{figure}
  \centering
  \includegraphics[width=0.45\textwidth]{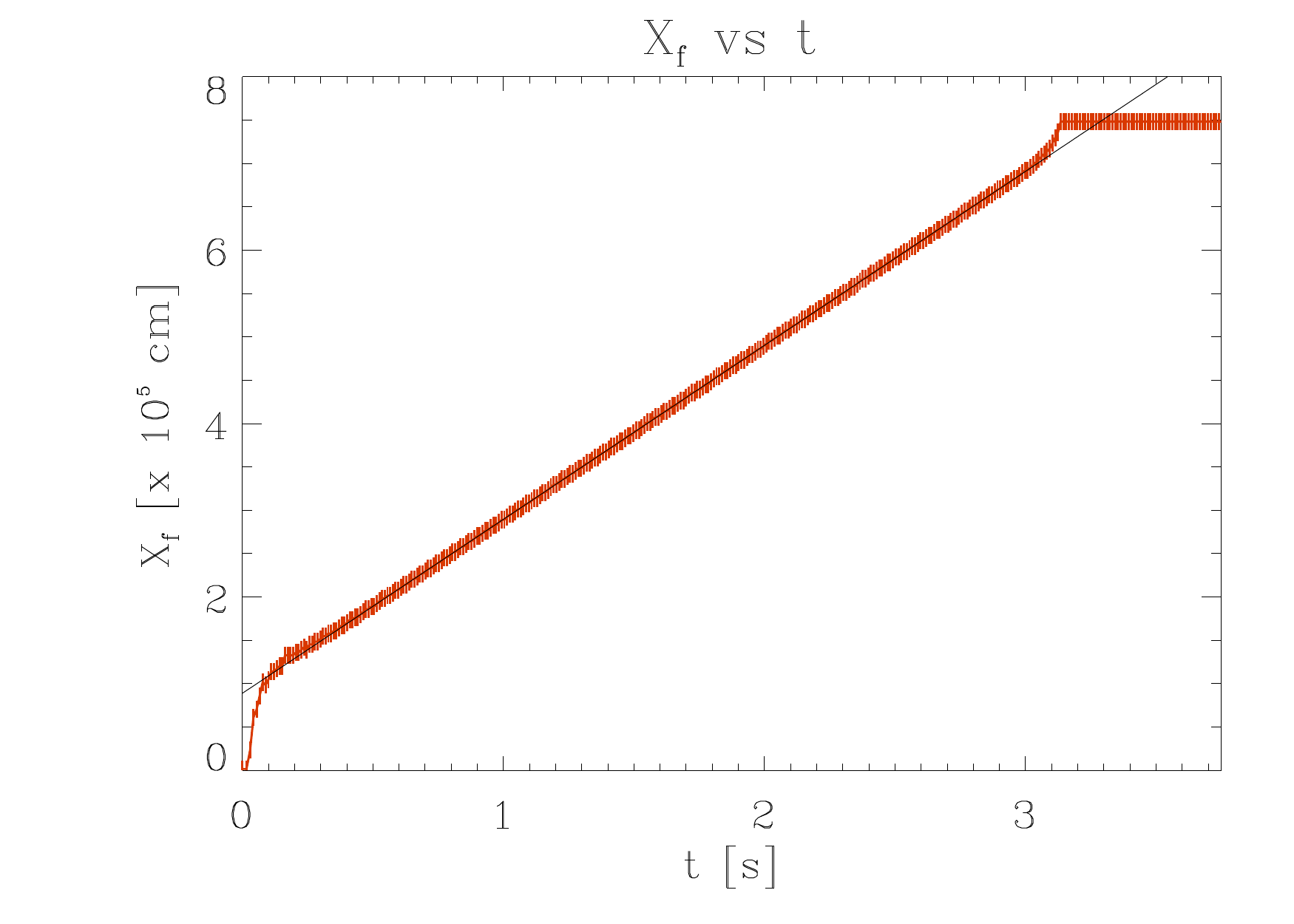}
  \caption{Flame front position for run 6.The symbols indicate the
    error bars on the positions, while the line shows the best linear
    least squares fit. After an initial stage, the flame adjusts to
    steady propagation. Eventually, the flame reaches the opposite
    boundary (in this case in $\sim 3$ s).}
  \label{fig:frontprop}
\end{figure}

To compute flame propagation speed from our simulations, we
define the front position as the location with the maximum burning
rate. In Fig. \ref{fig:frontprop}, we follow the position of the
front for simulation 6 and
plot it versus time. At the beginning there is a transitional stage
after the flame is started by the initial perturbation of the
temperature and the front adjusts to a steadily spreading
configuration (in $\lesssim 0.1$ s).  This steady propagation is well
fitted by a straight line, and the gradient gives us the speed of the
flame front. We repeat the fit procedure for all of the various runs: the
resulting front speeds $\vf$ are reported in Table \ref{tab:runs}.

Having measured front velocities, we can determine the effects of the
rotational spin $\Omega$ and the effective opacity $\kaco$ (a proxy
for the heat conductivity). Overall, the gradient of the lifted fluid
is steeper for higher $\Omega$, and the baroclinicity (the
misalignment between the iso--surfaces of density and pressure
measured by $\nabla P \times \nabla \rho$) along the interface tends
to increase with $\Omega$. The flame propagation speed decreases as
the rotation rate increases (see the next sections and Fig.
\ref{fig:vfomega}). Changing $\kaco$ also has an effect on flame
velocity: the flame is faster for lower $\kaco$, but the velocity
saturates when $\kaco \lesssim 10^{-2}$ cm$^2$ g$^{-1}$ (see Fig.
\ref{fig:vfkappa}).

In Appendix \ref{sec:convtest} we discuss the convergence rate of the
code. Even though the rate is lower than desirable, so that the actual
values of the flame speed should be slightly different in reality, we
can be confident that the general conclusions we draw are solid.  In
particular, increasing the resolution decreases the flame speed, which
is indicative of the fact that a detonation would not
develop. Moreover, the flame always reaches a steady state and the
structure of the front is as described above and in the following.

\subsection{A first set of conclusions}
\label{sec:firstset}

Although the fluid moves ageostrophically from behind the interface
forward, this motion does not go past the interface (Fig.
\ref{fig:anatomy1} bottom right). We interpret this as follows: the
fluid has expanded on the left of the front because of its higher
temperature, and the resulting horizontal pressure gradient pushes the
hot burning fluid to spill over the unburnt fluid. The Coriolis force,
however, diverts the horizontal $x$ motion into the horizontal $y$
direction and thus creates a geostrophic current that compensates for
the horizontal pressure gradient. The resulting configuration is that
of the Rossby adjustment problem \citepalias[see][]{\bc}, as
anticipated by \citetalias{art-2002-spit-levin-ush}. In this case the
inclination angle $\alpha$ of the interface should be $\alpha\sim
H/\ro$, where $H$ is the scale height of the fluid and $\ro$ is the
Rossby radius $\ro=\sqrt{g H}/2\Omega$ (where $\Omega=2\pi\nu$ and
$\nu$ is the spin frequency of the NS). Measuring the slope of the
black line in Fig. \ref{fig:advcond}, we find that the slope is
$\alpha\sim 3.5\tent{-3}$, so that its horizontal extent is $\sim
2-3\ro$.  This is in accordance with expectations.

Regarding the motion that we observe in the vertical plane along the
interface, we note that here the fluid is much more baroclinic, that
is to say, the iso--surfaces of density and pressure are much more
misaligned than elsewhere, as can be seen in the lower right panel of
\begin{figure}
  \centering
  \includegraphics[width=0.45\textwidth]{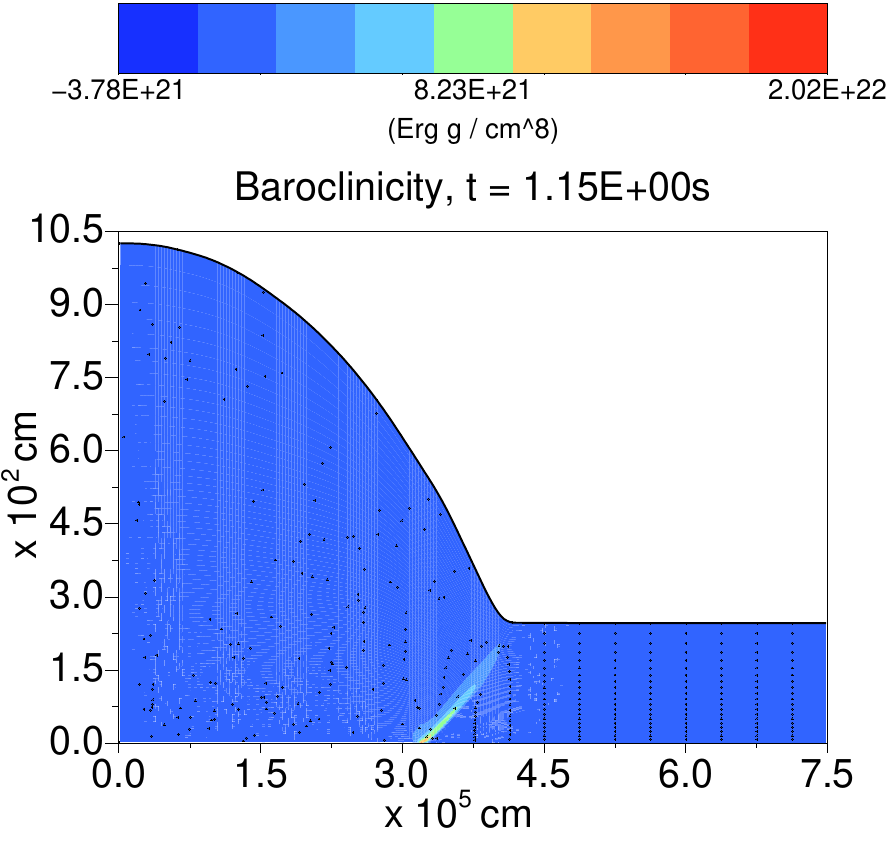}
  \caption{Baroclinicity: $\nabla P \times \nabla \rho$, for reference
    simulation 6. The vector is along the $y$ direction, coming out
    of the plane.}
  \label{fig:baroc}
\end{figure}
Fig. \ref{fig:anatomy1} and in Fig. \ref{fig:baroc}. It is well
known from geophysical studies that geostrophic balance is unstable in
the presence of baroclinicity. The resulting instability is similar in
nature to convection, but with motion, which is no longer vertical,
lying within the ``wedge of instability'' between the iso--surfaces of
pressure and density \citep{book-1987-Pedlo}.  \citet{art-1988-fuji,
  art-1993-fuji} and \citet{art-2000-cum-bild} in fact already studied
the possibility of baroclinic instability in the context of Type I
bursts, but their baroclinicity was very mild since they considered
the effects of shear induced by the differential rotation due to the
vertical expansion of the burning layer, and not the effects of the
huge horizontal temperature gradients that develop during flame
propagation.

\begin{figure}
  \centering
  \includegraphics[width=0.45\textwidth]{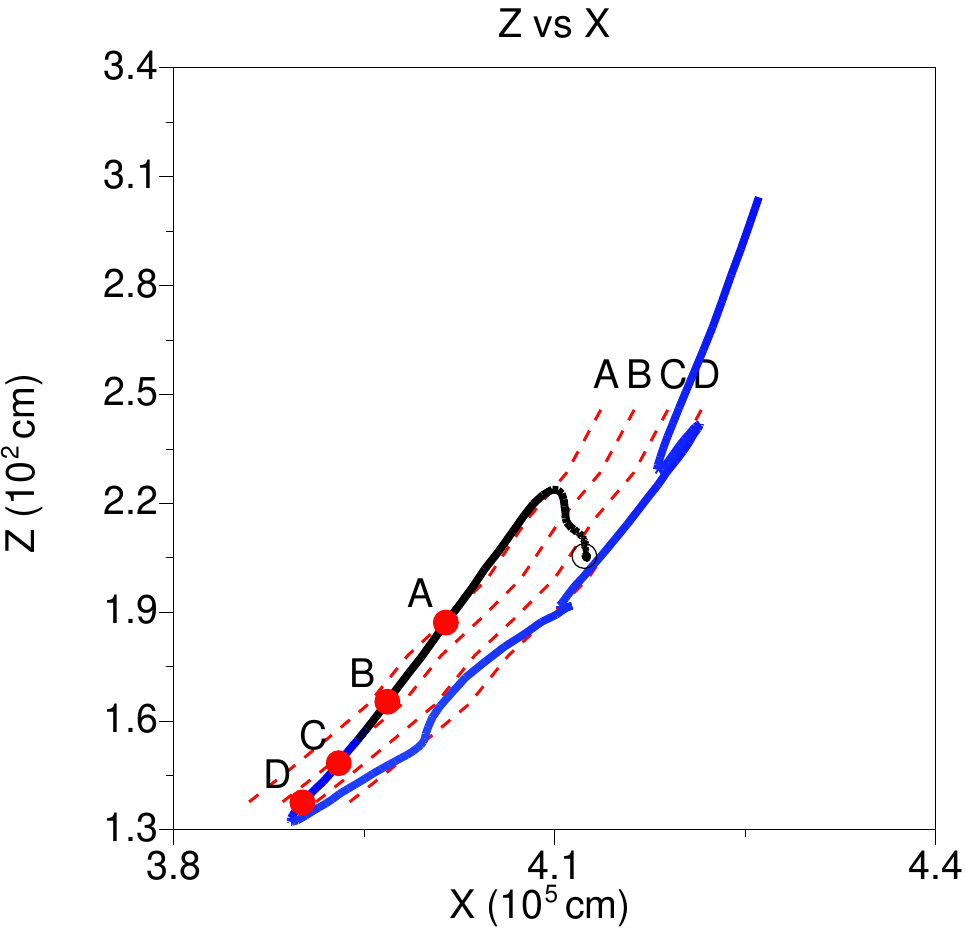}
  \caption{Example of motion of one tracer particle on the vertical
    plane. The red dashed lines indicate the position of the flame
    front at different times A, B, C and D. The corresponding
    positions of the particle are indicated by the same letters on the
    particle trajectory. The empty circle indicates the starting
    point. The colours of the trajectory correspond to the same scale
    as the contours for the burning rate in the middle panel of Fig.
    \ref{fig:anatomy1}.}
  \label{fig:particlemotion}
\end{figure}

In our case, the source of baroclinicity is the
horizontally-inhomogeneous nuclear burning\footnote{Compare Fig.
  \ref{fig:baroc} to the middle right panel of Fig.
  \ref{fig:anatomy1}.} which affects the temperature profile.  Its
steady propagation is maintained by the Coriolis force, which
reinforces the near-geostrophic configuration on time scales of the
order of $1/\nu$.  Following the tracer particle motion, we can see
advection along and in front of the interface, which we attribute to
the development of baroclinic instability.  In the previous section we
noted the presence of cells that are highly elongated in the
horizontal direction at the upper end of the hot-cold fluid interface:
we identify these cells with baroclinicity-induced motion. Fig.
\ref{fig:particlemotion} shows how particles are driven into the front
and down along the interface. After the flame has passed and the front
is farther away, the particles are caught by the advective motion and
driven upwards. The ascending part is different for different
particles and this picture just indicates the general trend.

\subsection{Front propagation mechanism}
\label{sec:frontmecha}

Summarizing the results from section \ref{sec:anatomy}, we see that at
the top of the interface the fluid is heated up by the spilling over
of the hot fluid, via advection and thermodynamics. However, in the most
relevant regions for flame propagation, heat is brought across the
interface primarily by conduction (mainly vertically given the small
inclination angle\footnote{Heat conduction in the horizontal direction
  can be neglected since the horizontal length scale is larger by a
  factor $\sim 10^3$ than the vertical length scale. Two runs
  where in one case full conduction was implemented and in the other
  only vertical conduction was used gave virtually identical results.}),
with a contribution from baroclinic instability mixing.

The contribution to $\partial T / \partial t$ in equation
\eqref{eq:progT} from conduction can be written by means of equation
\eqref{eq:phycond} as
\begin{equation}
\dtempdt{cond}=\frac{Q_{\rm{cond}}}{\tilde c_{\rm{P}}}=
\frac{1}{\tilde c_{\rm{P}}\rho}\nabla\cdot\left(
\frac{16}{3}\frac{\sigma_{\rm B} T^3}{\rho\kaco}\nabla T\right).
\end{equation}
From this we can derive an approximate diffusion coefficient for
conduction $D_{\rm cond}$ as
\begin{equation}
D_{\rm cond}\sim\frac{16\sigma_{\rm B} T^3 }{3 \tilde c_{\rm{P}}
\rho^2\kaco}
\end{equation}
and then derive the timescale for heat diffusion across the vertical scale
height $H$ as $\tau_{\rm cond}\sim H^2/D_{\rm{cond}}$, or
\begin{equation}
\label{eq:taucond2}
\tau_{\rm cond}\sim \frac{3}{16}\frac{\rho^2 H^2 \tilde c_{\rm{P}}}{\sigma_{\rm
    B} T^3}\kaco
\end{equation}
Equation \eqref{eq:taucond2} gives
\begin{multline}
\label{eq:taucond3}
\tau_{\rm cond}\sim
2.1\tent{-2}\;\rm{s}\;\scal{\kaco}{0.07\;\rm{cm}^2\;\rm{g}^{-1}}{}
\scal{\rho}{10^{5}\;\rm{g\; cm}^{-3}}{2}\\
\scal{T}{10^{9}\;\rm{K}}{-3}
\scal{H}{3\tent{2}\;\rm{cm}}{2}
\scal{\tilde c_{\rm{P}}}{10^8\;\rm{erg}\;\rm{K}^{-1}}{}
\end{multline}

Once the lower fluid has been heated up and starts burning it expands
again.  The Coriolis force then reinforces Rossby adjustment in a time
scale of the order of $\nu^{-1}\ll\tau_{\rm cond}$. This translates a
small vertical shift into a long horizontal displacement, where the
proportionality is given by the inclination of the interface:
$1/\alpha$ $\sim (2-3 \ro)/H$, as we will see in the next section.

The effective advective conduction brought about by baroclinic mixing
would act on a time scale given by
\begin{equation}
\label{eq:barotime}
\tau_{\rm bar}\sim H^2 / D_{\perp \rm{bar}},
\end{equation}
with
\begin{equation}
\label{eq:barodiff}
D_{\perp \rm{bar}}\sim w_{\perp \rm bar} \lambda_{\perp \rm bar},
\end{equation}
where $w_{\perp \rm bar}$ and $\lambda_{\perp \rm bar}$ are the
physical velocity of the fluid and its length scale perpendicular to
the hot-cold fluid interface.

As long as $\tau_{\rm bar} \gg \tau_{\rm cond}$, conduction will be
the most effective mechanism. 
\begin{table}
\centering
\begin{tabular}{|r|r|r|r|c|}
Run&$D_{\perp \rm{bar}}$&$\tau_{\rm{bar}}$&$\tau_{\rm{cond}}$&
$\tau_{\rm{bar}}/\tau_{\rm{cond}}$\\
\hline
 1$\phantom{^a}$&    223.0 &      403.6 &   142.5$\phantom{0}$   &   $\phantom{00}$2.8\\
 2$\phantom{^a}$&    288.6 &      311.8 &    99.8$\phantom{0}$   &   $\phantom{00}$3.1\\
 3$\phantom{^a}$&    328.9 &      273.6 &    71.3$\phantom{0}$   &   $\phantom{00}$3.8\\
 4$\phantom{^a}$&    449.5 &      200.2 &    42.8$\phantom{0}$   &   $\phantom{00}$4.7\\
 5$\phantom{^a}$&    915.0 &       98.4 &    14.3$\phantom{0}$   &   $\phantom{00}$6.9\\
 6$\phantom{^a}$&    926.0 &       97.2 &    10.0$\phantom{0}$   &   $\phantom{00}$9.7\\
 7$\phantom{^a}$&   1045.1 &       86.1 &     7.1$\phantom{0}$   &   $\phantom{0}$12.1\\
 8$\phantom{^a}$&   2047.2 &       44.0 &     1.4$\phantom{0}$   &   $\phantom{0}$31.4\\
 9$\phantom{^a}$&   7757.8 &       11.6 &     0.1$\phantom{0}$   &               116.0\\
10$^a$          &   ---    &        --- &     10.0$\phantom{0}$  &   $\phantom{00}$---\\
11$^a$          &   ---    &        --- &     10.0$\phantom{0}$  &   $\phantom{00}$---\\
12$\phantom{^a}$&    548.4 &      164.1 &    10.0$\phantom{0}$   &   $\phantom{0}$16.4\\
13$\phantom{^a}$&   1570.9 &       57.3 &    10.0$\phantom{0}$   &   $\phantom{00}$5.7\\
\hline
\end{tabular}
\caption{Diffusion coefficient for the baroclinicity driven advection 
  (equation \ref{eq:barodiff}) as measured directly from the simulations, 
  its diffusion timescale according to equation \eqref{eq:barotime} and the
  timescale for conduction as from equation \eqref{eq:taucond3}, using 
  $T=6\tent{8}$ K and $\rho=9\tent{5}$ g cm$^{-3}$. The last column reports 
  the ratio between the two timescales. Note that these 
  values are only indicative order of magnitude estimates.\newline $^a$ These runs
  had less clear configurations, so that reliable measurements were not possible.}
\label{tab:baroc}
\end{table}
In order to get a handle on the importance of the baroclinicity
induced advection, we measured the average values of $w_{\perp \rm
  bar}$ and $\lambda_{\perp \rm bar}$. For each grid point along the
interface, we calculated the component of fluid velocity in the
direction perpendicular to the interface. Considering only the region
over which $w_{\perp \rm bar}$ was negative, we calculated the average
$w_{\perp \rm bar}$ and measured $\lambda_{\perp \rm bar}$ as the
length of this region in the direction perpendicular to the
interface. We then calculated the total average $D_{\perp \rm{bar}}$:
the results are reported in Table \ref{tab:baroc} for all of the
simulations. The results confirm that baroclinicity is negligible most
of the time, apart from in the cases of very low heat diffusivity
(high $\kaco$). One should be aware, however, that all these values
are order of magnitude estimates, and hence only describe trends, not
precise timescales.

\subsection{Front propagation speed}
\label{sec:velmecha}

Following e.g. \citet{book-1959-landau-lif} and
\citet{art-1982-fryx-woos-b}, the velocity of flame propagation across
the interface should be given, in a deflagration regime, by
\begin{equation}
\label{eq:vperpflame}
v_{\rm{f}\perp}\sim\sqrt{\frac{D_{\rm cond}}{\tau_{\rm n}}}
\end{equation}
where $\tau_{\rm n}$ is the burning time scale, given by $\tau_{\rm
  n}=\epsilon_\alpha/Q_{\rm n}$, see Equations \eqref{eq:nukeburning}
and \eqref{eq:Yprogtot},
\begin{multline}
\tau_{\rm n}\sim1.1\tent{-1}Y^{-3}
\exp(4.4\tent{9}\;\rm{K}/T)\\
\scal{T}{10^9\;\rm{K}}{3}
\scal{\rho}{10^5\;\rm{g}\;\rm{cm}^{-3}}{-2}.
\end{multline}
In order to estimate the horizontal propagation velocity across the
NS, this velocity has to be multiplied by the factor $(2-3) \ro/H$
which expresses the ratio of the area of the burning front to the area
of the vertical section of the ocean
\citep[see][]{book-1959-landau-lif}.  The horizontal velocity becomes
\begin{equation}
\vf\sim\sqrt{\frac{16\sigma_{\rm B}}{3\tilde c_{\rm{P}}}
   \frac{ T^3}{\rho^2\tau_{\rm n}}\frac{g}{H}}\frac{1}{2\pi\nu\sqrt{\kaco}}
\end{equation}
or
\begin{multline}
\label{eq:expvf}
\vf\sim1.8\tent{6}Y^{3/2}\exp(-2.2\tent{9}\;\rm{K}/T)
    \;{\rm {cm\; s}^{-1}}\\
    \scal{\nu}{450\; \rm{Hz}}{-1}
    \scal{\kaco}{0.07\;\rm{cm}^2\; \rm{ g}^{-1}}{-1/2}
    \scal{g}{2\tent{14}\;\rm{cm\;s}^{-2}}{1/2}\\
    \scal{H}{3\tent{2}\;\rm{cm}}{-1/2}
    \scal{\tilde c_{\rm P}}{10^8\;\rm{erg\;g^{-1}\;K^{-1}}}{-1/2}
\end{multline}

Again, if $\tau_{\rm bar} \sim \tau_{\rm cond}$, then the actual
$v_{\rm{f}\perp}$ should be given by a combination of conduction and
advection, with an extra term of the order of
\begin{equation}
\sqrt{\frac{D_{\perp \rm bar}}{\tau_{\rm n}}}.
\end{equation}
to be included in equation \eqref{eq:vperpflame}. The result should
then be multiplied by the same factor, $(2-3)\ro/H$.

The scaling expected from equation \eqref{eq:expvf} agrees with what
we measure in our simulations. The agreement is within half an order of
magnitude and this allows us to put constraints on the numerical
factors in front of equation \eqref{eq:expvf} which are not determined
by the order of magnitude estimates that led to it.

\begin{figure}
  \centering
  \includegraphics[width=0.45\textwidth]{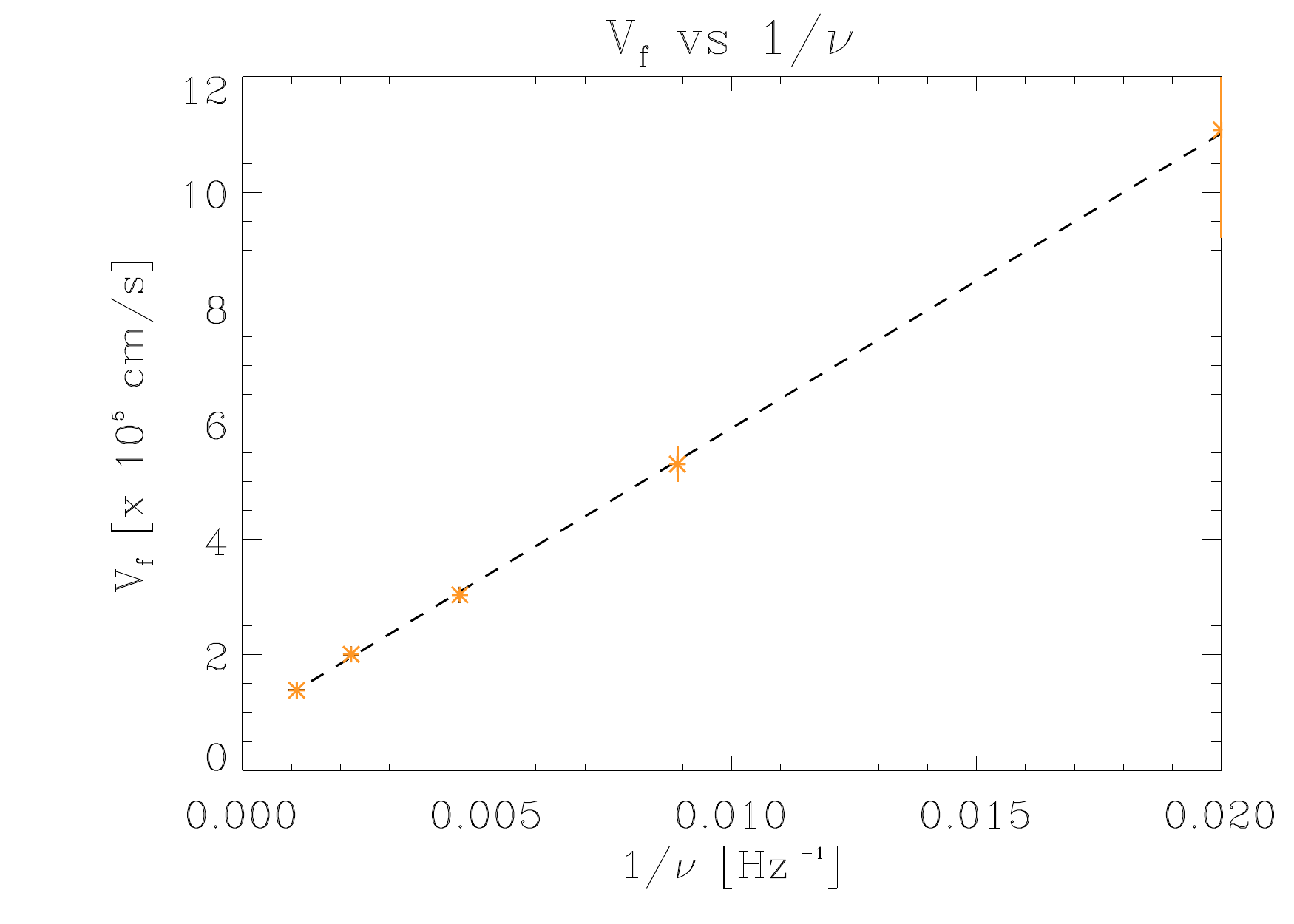}\\
  \includegraphics[width=0.45\textwidth]{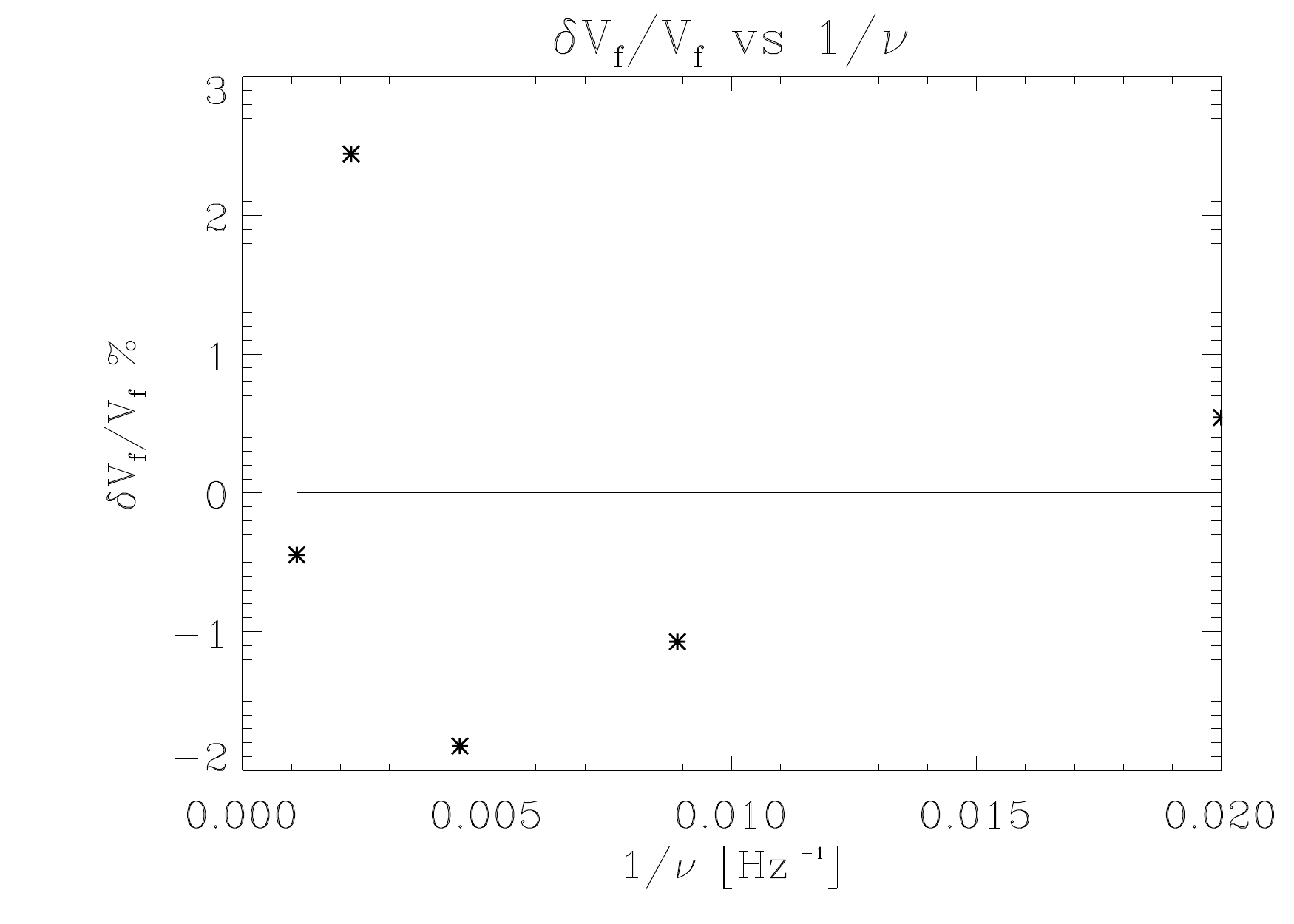}
  \caption{Upper panel: velocity of flame propagation versus
    $1/\nu$. Lower panel: residuals with respect to the best fit
    through the points versus $1/\nu$. All these runs have
    $\kaco=0.07$ cm$^2$ g$^{-1}$.}
  \label{fig:vfomega}
\end{figure}

We verify the dependence of $\vf$ on the spin rate $\nu$ using runs
6-10. In Fig. \ref{fig:vfomega}, we
can see that increasing the rotation frequency slows down the flame,
as expected, with a $1/\nu$ dependence. The dotted line in the upper
panel shows the best linear fit to the data, which has a slope of
$5.10\tent{7}$ cm s$^{-1}$ and an intercept of $8.25\tent{4}$ cm
s$^{-1}$. The lower panel shows the relative difference between the
fit and the results from the simulations.

The presence of the intercept at $1/\nu=0$ is the most notable
feature. This intercept is not predicted by the back of the envelope
calculations leading to equation \eqref{eq:expvf}. Nonetheless, it is
to be expected physically that even in the presence of extremely fast
rotation, which would lead to a \emph{vertical} interface, there
should still be some conduction across the interface leading to a
finite front speed.

\begin{figure}
  \centering
  \includegraphics[width=0.45\textwidth]{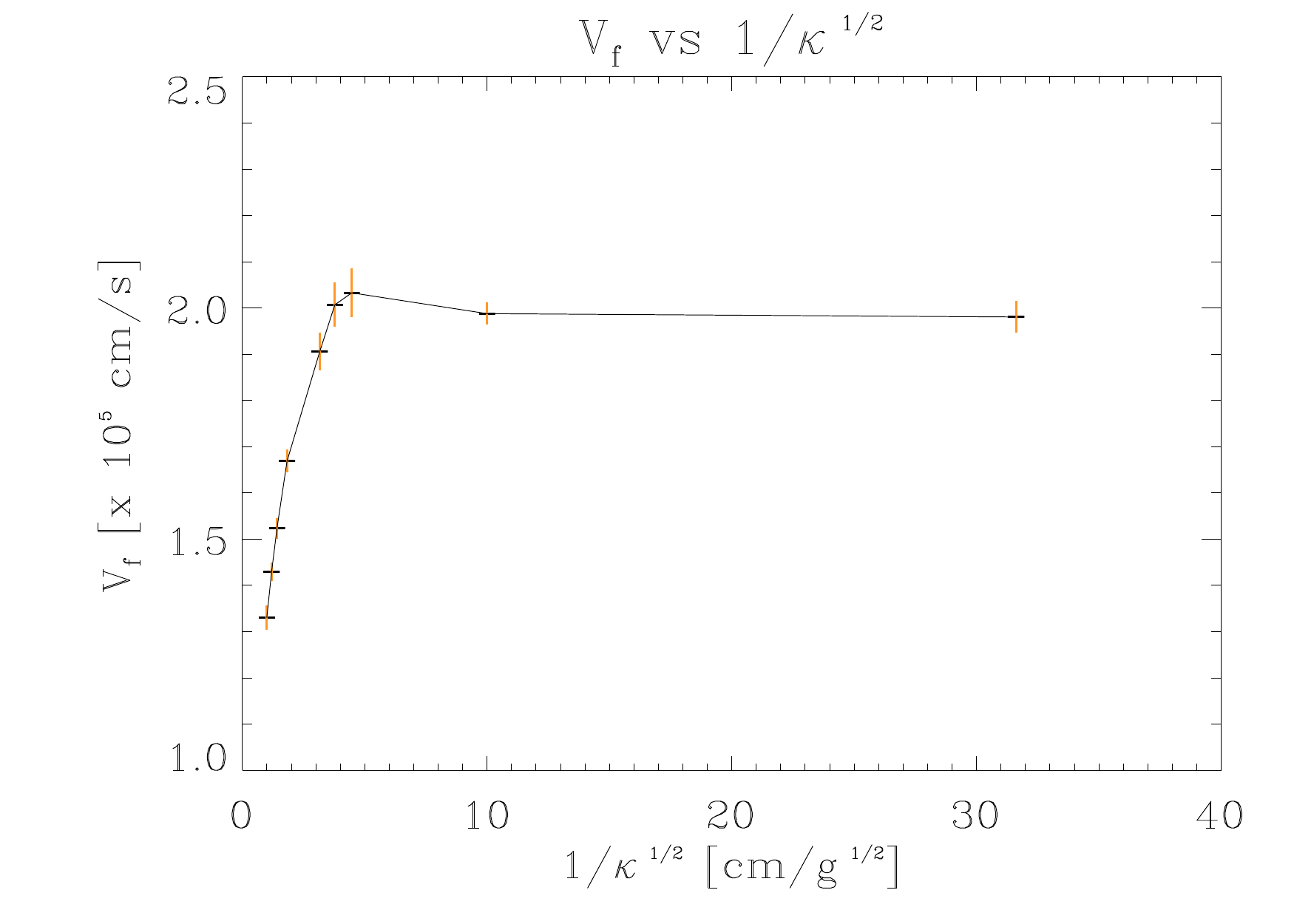}
  \caption{Velocity of flame propagation versus $1/\sqrt{\kaco}$.
    All these runs have $\nu=450$ Hz.}
  \label{fig:vfkappa}
\end{figure}

By contrast, Fig.
\ref{fig:vfkappa}, where we plot $\vf$ against the inverse heat
conduction (our
effective opacity $\kaco$), shows more complex behaviour. For
opacities $\kaco \gtrsim 0.05$ g cm$^{-2}$, the flame speed increases
approximately with $1/\sqrt{\kaco}$ as expected from equation
\eqref{eq:expvf}. Below it, the velocity seems to asymptote to $\vf
\sim 1.99\tent{5}$ cm s$^{-1}$. In our simulations we see a change in
the morphology of the flame: indeed all simulations showed a flame
\begin{figure}
  \centering
  \includegraphics[width=0.44\textwidth]{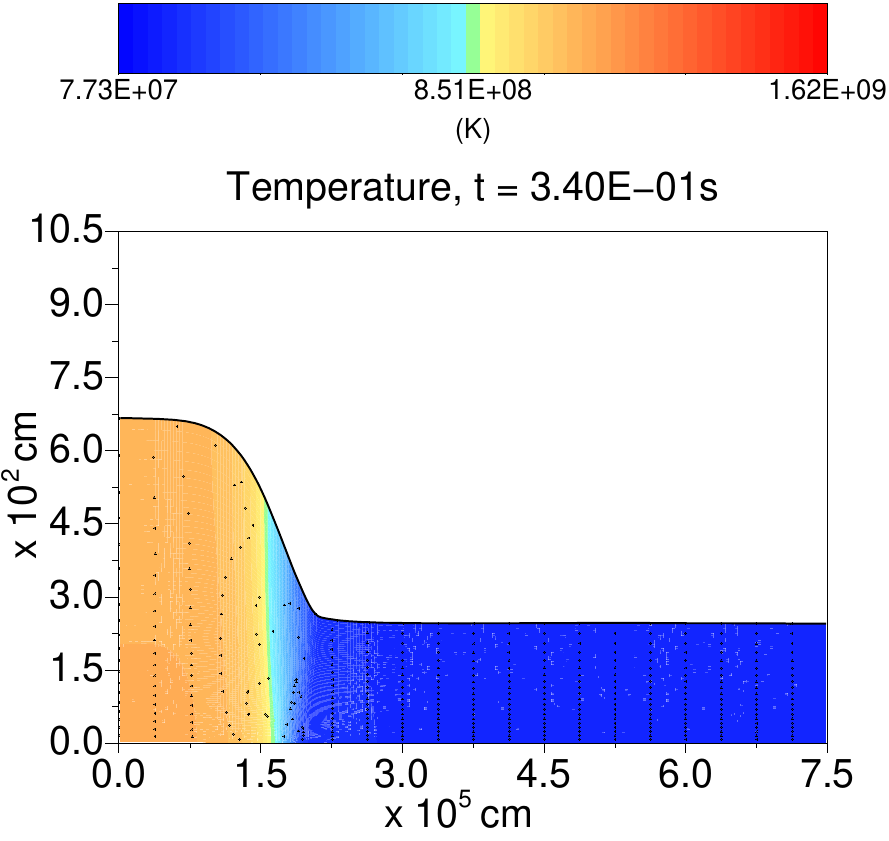}
  \includegraphics[width=0.44\textwidth]{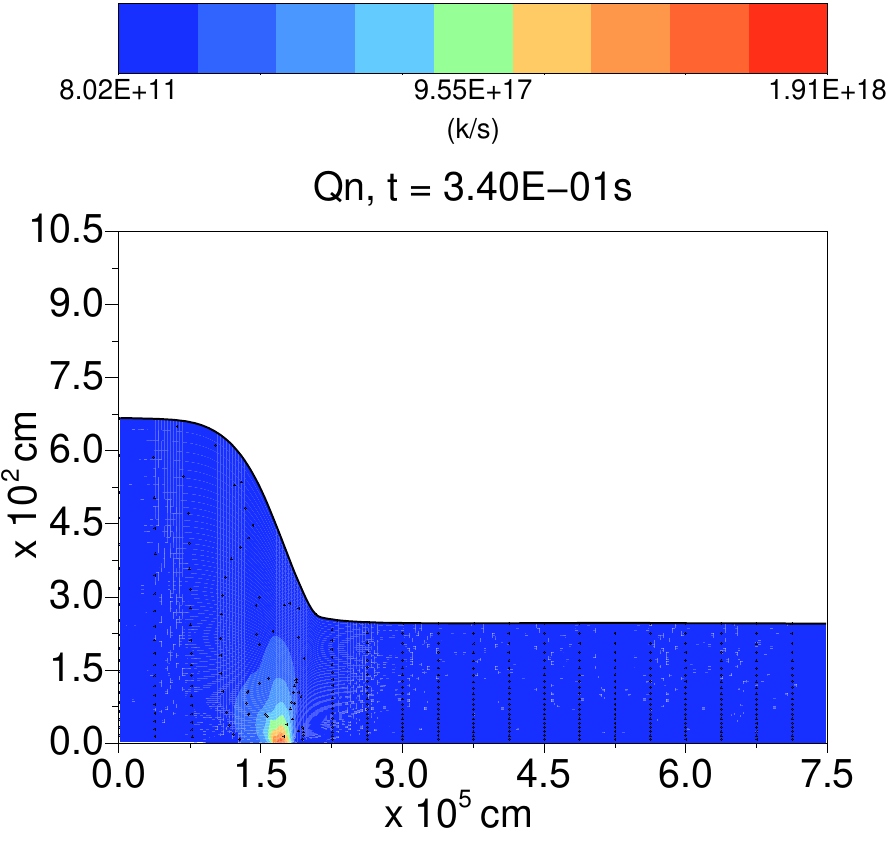}
  \caption{Temperature and reaction rate for simulation (9) after the
    flame is steadily propagating. The morphology of the flame is
    different from that shown in Fig. \ref{fig:anatomy1}. This
    simulation shows the asymptotic behaviour seen in Fig.
    \ref{fig:vfkappa}.}
  \label{fig:flame27}
\end{figure}
leaning on the hot-cold fluid interface similar to the middle right
panel of Fig. \ref{fig:anatomy1}, apart from simulation (9)
($\kaco=0.001$ g cm$^{-2}$), which does not show such a leaning flame
and has a much more vertical structure of the temperature profile
(Fig. \ref{fig:flame27}), and simulation (8) ($\kaco=0.01$ g
cm$^{-2}$) where this trend is beginning to become apparent.

We interpret this point as marking the transition where the conduction
time scale $\tau_{\rm cond}$ becomes comparable to the burning time
scale $\tau_{\rm n}$ and the thickness of the slanted burning front
becomes comparable to the vertical scale height. At this point the
front speed saturates and the whole layer burns through on the
timescale $\tau_{\rm{n}}$: according to the original estimate of
\citetalias{art-2002-spit-levin-ush} (see the discussion leading to
equations 21-23), this means that the horizontal speed saturates at
$\sim \ro/\tau_{\rm{n}}$. If we adopt the values $T=6\tent{8}$ K and
$\rho=9\tent{5}$ g cm$^{-3}$, as in Table \ref{tab:baroc}, we obtain
an \emph{average} burning rate of $\tau_{\rm{n}}\sim 0.45$ s, so that
$2\;\ro/\tau_{\rm{n}}\sim 1.9\tent{5}$ cm s$^{-1}$. This
order-of-magnitude estimate is in good agreement with what we measure.

On the other hand, when $\kaco\gtrsim 0.3$ g cm$^{-2}$, we observe a
greater deviation from our estimates.  In this limit of small
conductivity, we suspect that the baroclinic motions become more
important.  Their contributions, on top of those predicted by equation
\eqref{eq:expvf}, have to be taken into account, until the front speed
asymptotes to a baroclinic-motion-driven system.

\section{Discussion and conclusions}
\label{sec:conclusions}
 
In this paper, we have been able to simulate for the first time the
lateral propagation of a \emph{deflagrating} vertically resolved flame
on the surface of an NS.  We find that after an initial post-ignition
adjustment, the front propagates steadily with constant velocity,
until it reaches the opposite side of the simulation box. The fact
that the flame velocity is constant (Fig.  \ref{fig:frontprop}) gives
us confidence that, regardless of the physics of the localized
ignition, steady flame propagation depends only on the physics acting
in the ocean layer and the conditions therein.  After all the surface
has been traversed by the flame, the fluid column cools down slowly,
in a time which depends on the opacity (see Equation
\ref{eq:coolings}), whilst still burning the residual fuel.  We note
that in 2003, Anatoly Spitkovsky (unpublished) obtained somewhat
similar flame fronts using the \code{pencil} code. Due to
computational constraints, however, he assumed unphysically large NS
spins, so that the Rossby radius was comparable to the ocean scale
height.  The micro-physics of the flame propagation mechanism was not
identified, and full exploration of the parameter range was not
carried out (Spitkovsky, private communication).

We have explored the dependence of the flame speed on the spin
frequency of the NS $\nu$ and the heat conductivity of the fluid
(expressed as an inverse of the effective opacity $\kaco$). We
measured velocities in the range $1.33\tent{5}$ - $1.11\tent{6}$ cm
s$^{-1}$, which cross the entire domain of $7.5$ km in $0.7$ - $5.6$
s. These numbers are in good agreement with the rise times observed
from Type I burst sources, suggesting that we have included all the
relevant physics in our simulations and that we are now in a position
to explore in more detail the behaviour of flame propagation during
Type I bursts. We caution the reader again from taking the speed
values to be exact, given that our convergence tests suggest a
somewhat slow convergence rate so that the true values will be
slightly different; however, the conclusions are solid, especially the
ones about the physical mechanism of flame spreading.

The flame propagates through a combination of the ageostrophic forward
flow of the burning fluid on top of the as-yet unburnt fluid
\citepalias[as argued previously in][]{art-2002-spit-levin-ush}, and
top-to-bottom heat transport across the large-area strongly-inclined
interface between burning and cold fluid. Heat transport leading to
ignition is affected primarily by microscopic heat conduction and, in
runs where conductivity was set to lower values, by baroclinic
motions.

In section \ref{sec:velmecha} we derived an order of magnitude
estimate for the velocity that the front would have if it were driven
by conduction. We calculated a dependence of the speed on $\kaco$ and
$\nu$ of the form $1/\nu\sqrt{\kaco}$ (Fig.s \ref{fig:vfomega} and
\ref{fig:vfkappa}) and confirmed these expectations with the results
of our simulations. A breakdown of this $\kaco$ dependence is seen at
both low and high $\kaco$, which can be understood qualitatively. In
particular, we observe the existence of a possible asymptote in the
velocity when the effective opacity is too small, which we explain as
follows.  When the opacity decreases sufficiently, the conduction time
scale becomes shorter than the nuclear burning time scale.  The latter
becomes the bottleneck, the burning front width becomes comparable to
the scale height, and the nuclear burning time scale becomes the time
scale of vertical expansion. This translates into a horizontal
velocity of $\sim \ro/\tau_{\rm n}$, as already anticipated by
\citetalias{art-2002-spit-levin-ush}.

There are a number of hydrodynamical issues that now have to be
explored further. Firstly, the effect of the baroclinic instability at
the hot--cold fluid interface could be explored in more
detail. Secondly, the flow in the $y$-direction has a velocity
comparable to the sound speed, and Kelvin-Helmholtz instabilities that
might be generated by this flow need to be investigated.  Other
aspects of the flame propagation will be explored in future work,
including the effects of a better burning prescription taking into
account elements other than Helium.  We also aim to investigate the
possibility of exciting large-scale waves in the ocean, and the effect
of magnetic fields.

Finally, some of our simulations suggest that in the absence of a
sufficiently strong Coriolis force the flame will die out. This leads
to an important question: can the flame cross the equator?  Near the
equatorial belt the effective Coriolis force is much weaker and this
could lead to rapid lateral spreading of the burning front, and
subsequent quenching of the burning by enhanced cooling. This would
have important consequences for efforts to determine the NS radius
from observations of type-I bursts \citep[see,
e.g.,][]{art-2010-stei-lat-brow}, since it is usually assumed that the
whole star is burning at the peak, and the derived radius of the
burning area is used as a measure of the NS radius. If the flame
cannot cross the equator, this fact has to be taken into account when
dealing with those estimates.  This would also have important
implications for burst recurrence times, and may help to explain the
properties of multi-peak bursts \citep{art-2006-bhatta-stro}.  We plan
to investigate this possibility by introducing a variable Coriolis
parameter in future work, to simulate properly the changes that would
occur as a flame approaches the equatorial belt.
\\

{\it Acknowledgements.} We thank Frank Timmes for making his
astrophysical routines publicly available. We also thank Anatoly
Spitkovsky, Chris Matzner, Alexander Heger and Laurens Keek for useful
discussions. We thank Simon Portegies Zwart for letting us use the LGM
cluster (NWO grant no. 612.071.503) and Jeroen B\'edorf for his
help. This research was supported by NOVA and by internal grants from
Leiden Observatory. Some of the research was carried out during an
extended visit by YC to the School of Physics at Monash University,
and he thanks the School for hospitality.

%thebibliography
\bibliographystyle{mn2e}
\bibliography{ms}
\label{lastpage}

\appendix

\section{Convergence tests}
\label{sec:convtest}

In this appendix we show some of the convergence tests that were performed. We used
resolutions of 60x24, 120x48, 240x96 and 480x192 to test the numerical
properties of the code. We used the same values for the parameters as
for reference run 6 (see Table \ref{tab:runs}).

\begin{figure}
  \centering
  \includegraphics[width=0.43\textwidth]{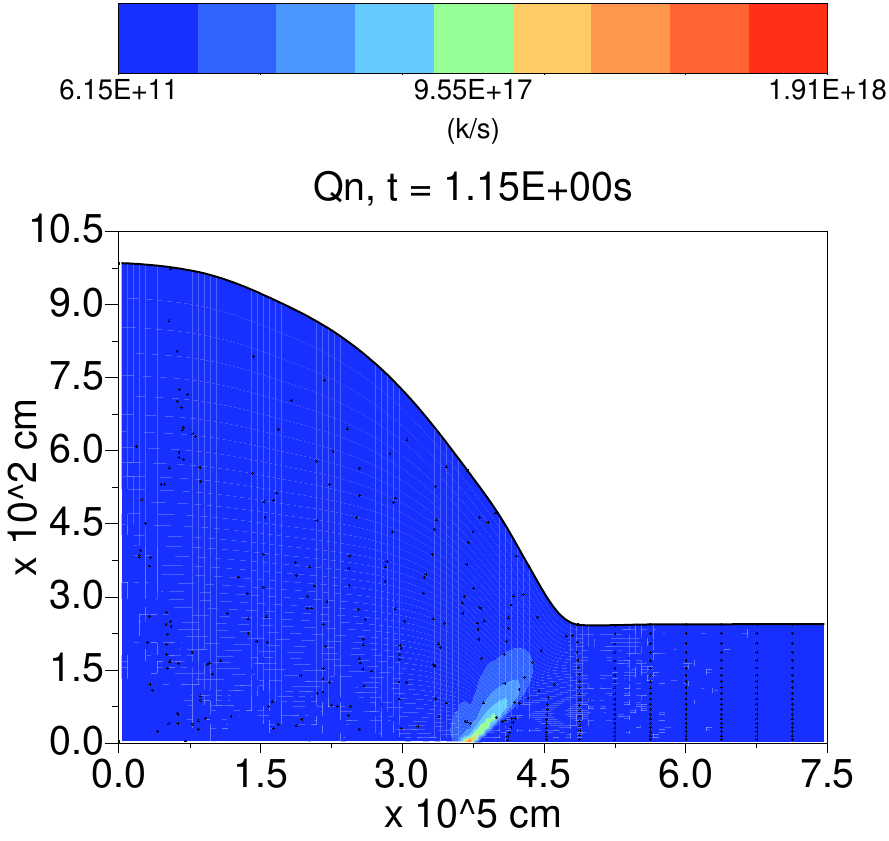}\\
  \includegraphics[width=0.43\textwidth]{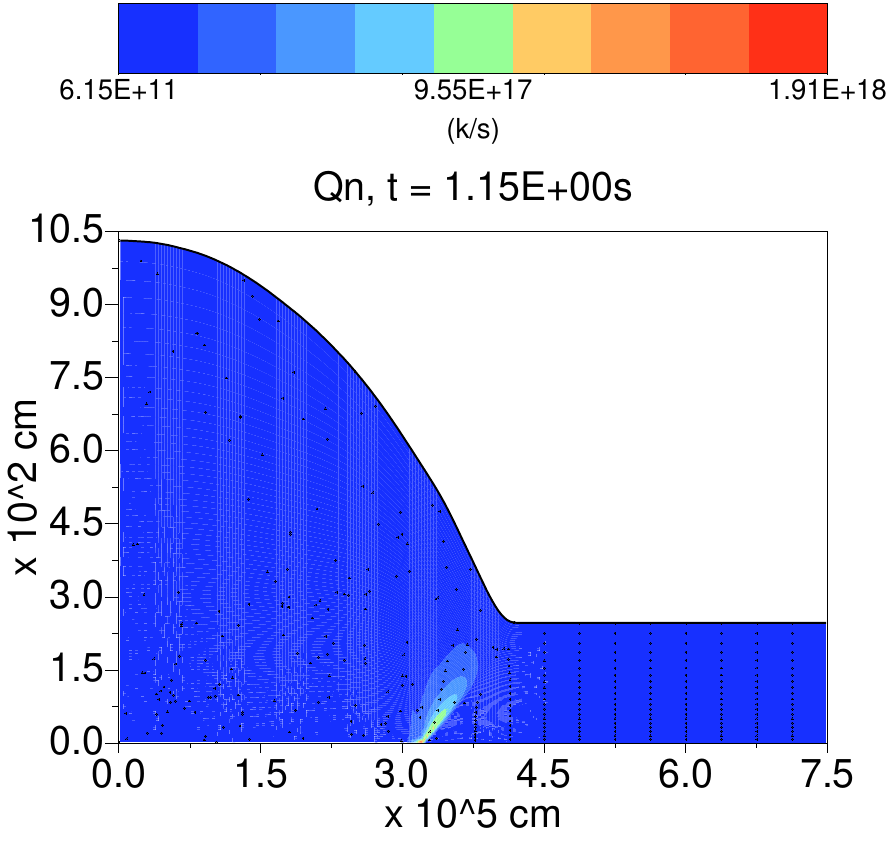}\\
  \includegraphics[width=0.43\textwidth]{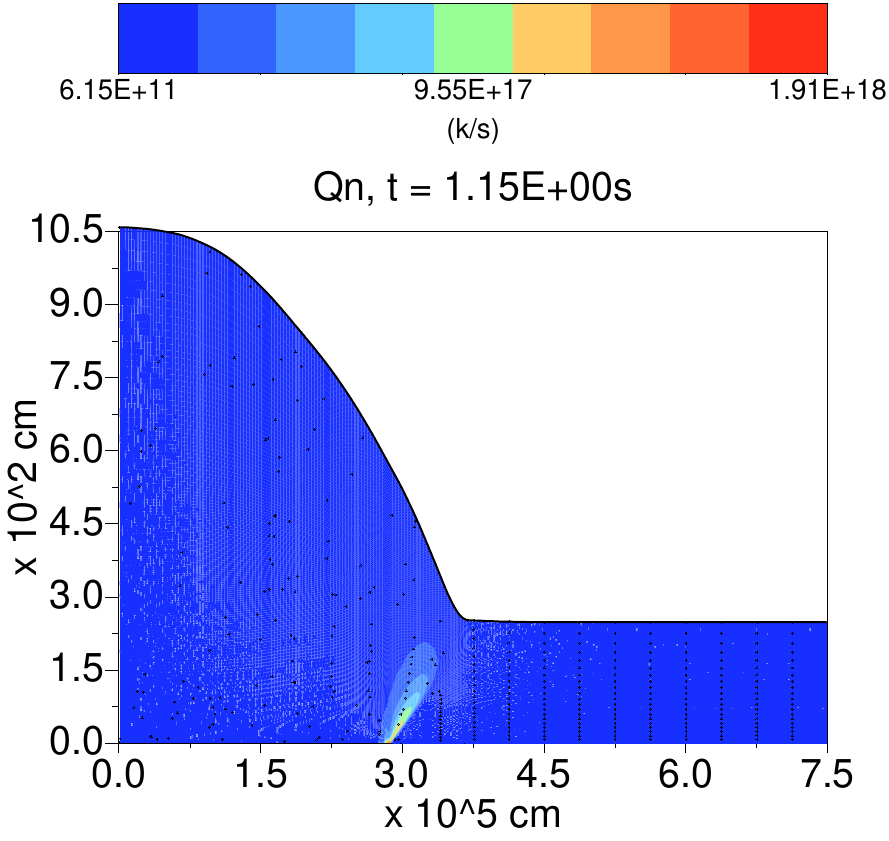}
  \caption{Snapshots of the burning rate for the simulations with
    resolutions 120x48, 240x96 and 480x192 at approximately the same
    time. The overall structure is the same, but the position of the
    flame is different due to the different values of the propagation
    speed. The different height of the fluid is just an artefact of
    the representation due to the different resolutions: we plot the
    centres of the grid cells and the higher the resolution the closer
    the centre of the top cell is to the physical top of the
    simulation.}
\label{fig:convflame}
\end{figure}

First of all, we notice that the behaviour at different resolutions is
qualitatively the same, showing the same initial transient phase and
then stationary propagation (see Fig. \ref{fig:convflame}), with
particles moving along the interface due to the increased
baroclinicity.

\begin{figure}
  \centering
  \includegraphics[width=0.45\textwidth]{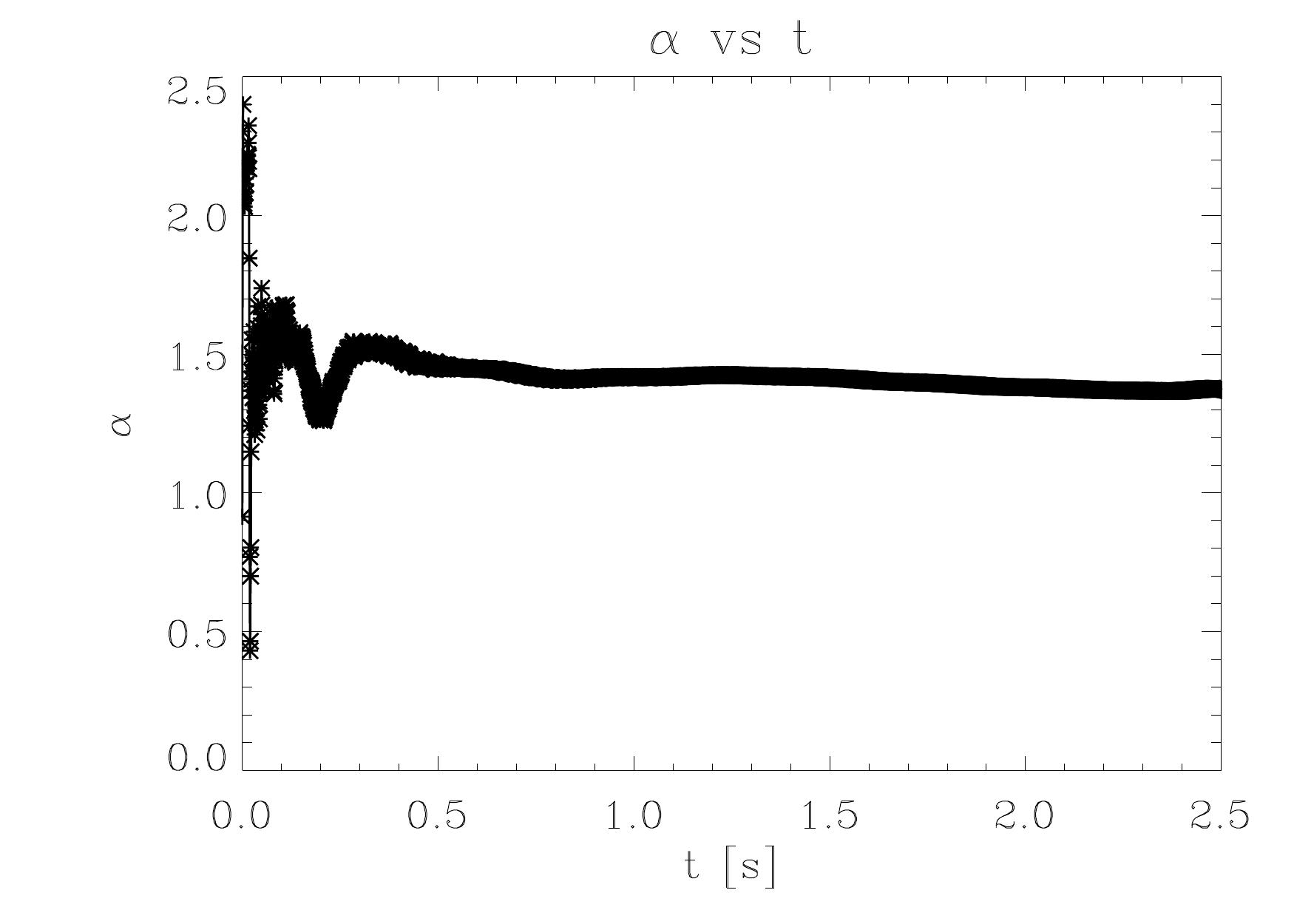}
  \caption{Convergence rate $\alpha$ (equation \ref{eqalpharate}) for
    temperature as a function of time for the simulations with
    resolutions 240x96, 120x48 and 480x192. The convergence rate is
    hampered by the different velocities of the flame front. The
    comparison is carried out until the flame is present in all
    simulations.}
  \label{fig:convrate}
\end{figure}

Secondly, we measure the convergence rate $\alpha$ according to
\begin{equation}
  \alpha=\frac{\sum_{i,j}\left|T_{2\,i,j} - T_{1\,i,j} \right|}
  {\sum_{i,j}\left|T_{1\,i,j} - T_{0.5\,i,j} \right|},
\label{eqalpharate}
\end{equation}
where the subscripts $2$, $1$, and $0.5$ refer to the simulations with
120x48, 240x96 and 480x192; $i\in[1,120]$ and $j\in[1,48]$. The values
of $T$ for the higher resolution simulations are interpolated at the
corresponding positions for the lowest resolution simulation. Since
the timesteps are slightly different, we also needed to perform a
linear interpolation in time. The results as a function of time are
plotted in Fig. \ref{fig:convrate}. This kind of convergence test is
hampered by the fact that the flame speed is different and this gives
high gradients at different spatial locations. Also, in increasing the
resolution we are actually simulating different physics, since, for
example, the convection that we can resolve is different: how to
predict the effect of that on $\alpha$ is difficult and beyond the
scope of this appendix. However, the fact that resolving more the
convection cells has an effect on the convergence rate becomes
apparent if we separate the domain in three horizontal domains: one
from the left boundary up to the beginning of the flame (of the
highest resolution simulation), one from the front of the flame (of
the lowest resolution simulation) to the right boundary and one
between these two. The first one, which is where we see the vertical
convective cells, has a convergence rate lower than the average, while
the second has a much better convergence rate. The middle one, which
by construction does not encompass only the flame, has a convergence
rate similar to the average one. The convergence rate using the
simulations with 60x24, 120x48 and 240x96 is only slightly better.

\begin{figure}
  \centering
  \includegraphics[width=0.45\textwidth]{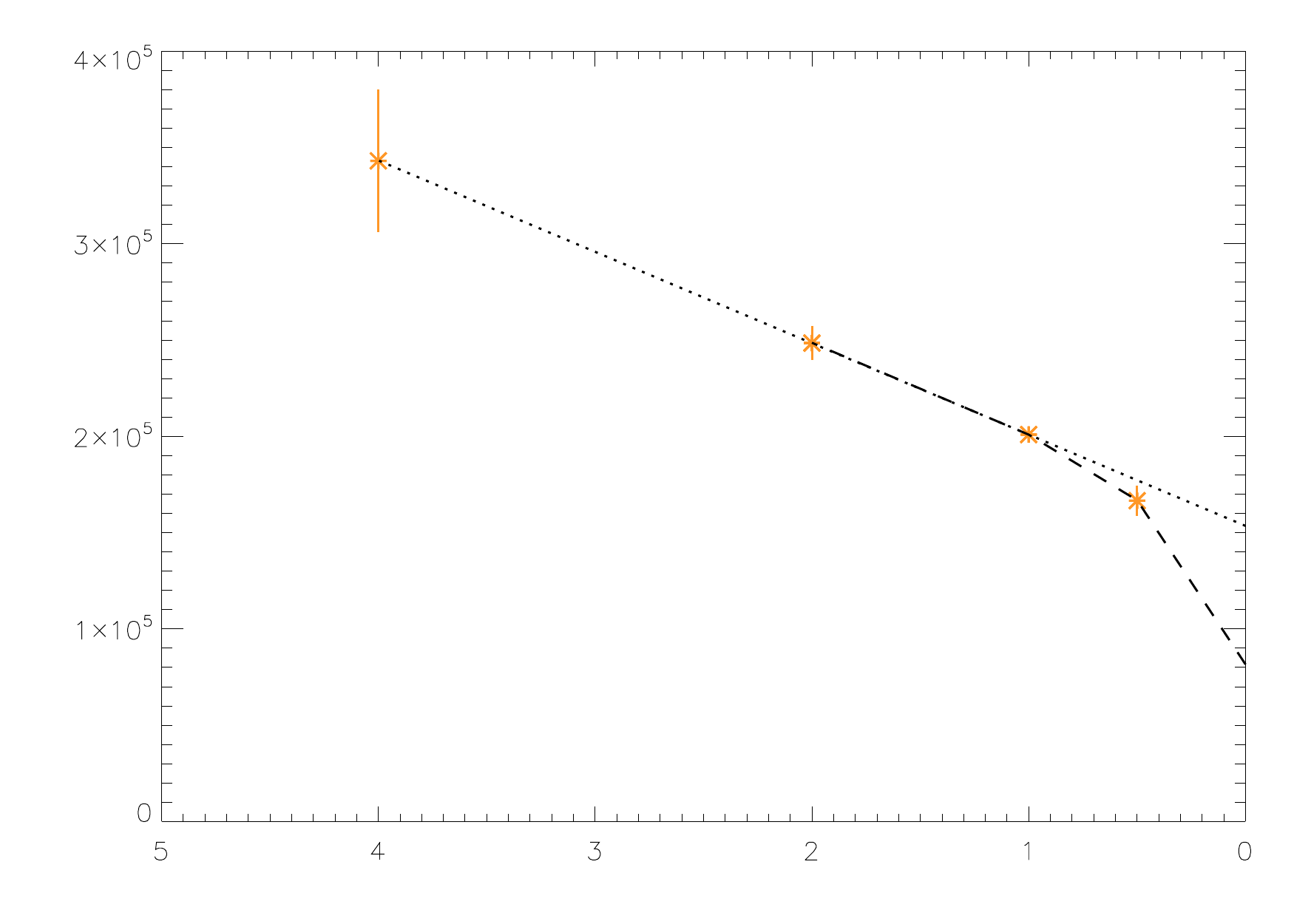}
  \caption{Flame spreading velocities as a function of grid spacing
    (simulation with 240x96 is the reference one). The dotted line
    indicates the linear fit and the dashed curve the non linear
    one. In both cases the velocity appears to be converging to a
    non-zero value, of at least $8.16\tent{4}$ cm s$^{-1}$.}
  \label{fig:convvelocities}
\end{figure}

If we measure a global quantity such as the velocity of the front, the
results are: $3.43\tent{5}$, $2.48\tent{5}$, $2.01\tent{5}$ and
$1.67\tent{5}$ cm s$^{-1}$. The convergence rate for the first three
values is $\alpha=2$, which basically implies linear convergence with
the resolution, while it becomes $\alpha=1.4$ when we consider the
three simulations with higher resolution. If we were to fit a line
through the first three values, the extrapolation for the ideal
infinite resolution would be $1.54\tent{5}$ cm s$^{-1}$ (see Fig.
\ref{fig:convvelocities}), while if we were to fit a non linear
function of the kind $v=a + b h^\beta$, with $\beta=\log 1.4 / \log
2$, then the extrapolation would be $8.16\tent{4}$ cm s$^{-1}$, which
is still non-zero (Fig. \ref{fig:convvelocities}). The real expected
value for an infinitely resolved simulation should lie between those
two. Finally, one \emph{very} important aspect to point out is that
the speed of the flame is \emph{decreasing} with \emph{increasing}
resolution. This is a very good sign that the motion of particles we
see should not develop into turbulence, hence possibly triggering a
detonation.

Finally, measuring the independent residuals
\begin{equation}
  I =\frac{ \sum_{i,j}\left|\frac
    {\left[T_{i,j}(t+dt) - T_{i,j}(t)\right]/dt - \partial T_{i,j} / \partial t}
      {\partial T_{i,j} / \partial t}\right|}
  {mx \, mz}
\end{equation}
where $mx$ and $mz$ are the resolutions in the horizontal and vertical
directions, gives results that are at most $3\tent{-8}$, so that
they are never a problem.

%end of document
\end{document}